\documentclass{article}
\usepackage{arxiv}

\usepackage[utf8]{inputenc} 
\usepackage[T1]{fontenc}    
\usepackage{hyperref}       
\usepackage{url}            
\usepackage{booktabs}       
\usepackage{amsfonts}       
\usepackage{nicefrac}       
\usepackage{microtype}      
\usepackage{lipsum}		
\usepackage{graphicx}
\usepackage{doi}

\usepackage{amssymb}
\usepackage{amsthm}
\usepackage{amsmath}
\usepackage{lineno}
\usepackage{comment}
\usepackage{lscape}
\usepackage{float}
\usepackage{caption}
\usepackage{todonotes}
\usepackage{soul}
\usepackage{multirow}
\usepackage{multicol}

\usepackage{makecell}
\setcellgapes{5pt}

\title{The AI Mechanic: Acoustic Vehicle Characterization Neural Networks}

\author{ Adam M. Terwilliger \qquad Joshua E. Siegel \\
	Department of Computer Science and Engineering\\
	Michigan State University\\
	East Lansing, MI, 48824 \\
	\texttt{adamtwig@msu.edu, jsiegel@msu.edu} \\
}

\renewcommand{\shorttitle}{Acoustic AI Mechanic}

\begin{document}
\maketitle

\begin{abstract}
In a world increasingly dependent on road-based transportation, it is essential to understand vehicles. We introduce the AI mechanic, an acoustic vehicle characterization deep learning system, as an integrated approach using sound captured from mobile devices to enhance transparency and understanding of vehicles and their condition for non-expert users. We develop and implement novel cascading architectures for vehicle understanding, which we define as sequential, conditional, multi-level networks that process raw audio to extract highly-granular insights. To showcase the viability of cascading architectures, we build a multi-task convolutional neural network that predicts and cascades vehicle attributes to enhance fault detection. We train and test these models on a synthesized dataset reflecting more than 40 hours of augmented audio and achieve $>92\%$ validation set accuracy on attributes (fuel type, engine configuration, cylinder count and aspiration type). Our cascading architecture additionally achieved $93.6\%$ validation and $86.8\%$ test set accuracy on misfire fault prediction, demonstrating margins of $16.4\%$ / $7.8\%$ and $4.2\%$ / $1.5\%$ improvement over na{\"i}ve and parallel baselines. We explore experimental studies focused on acoustic features, data augmentation, feature fusion, and data reliability. Finally, we conclude with a discussion of broader implications, future directions, and application areas for this work. 
\end{abstract}

\keywords{Emerging applications and technology \and Audio classification \and Acoustic classification \and Sound recognition \and Vehicle attributes \and Engine misfire detection \and Fault detection \and Artificial intelligence \and Deep learning \and Neural networks \and Cascading architecture \and Conditional architecture \and FFT \and MFCC \and Spectrogram \and Wavelets}

\section{Introduction}

\label{sec:intro}
Transportation is essential to daily life, with over one billion automobiles in use globally~\cite{voelcker_2014}. These vehicles produce approximately three billion metric tons of carbon dioxide annually~\cite{tiseo_2021} and consume scarce natural resources. As miles travelled increase, the need to reduce emissions and conserve energy will grow in importance.\\

Further, increases in autonomy and shared mobility will soon demand that expensive highly-automated vehicles be amortized over more miles travelled~\cite{sae2018definitions, bimbraw2015autonomous, bagloee2016autonomous}. A growing market for personal automobiles, alongside increased longevity requirements, decreased supply, and demand for long-life vehicles such as those to be used in new mobility services has led to more demanding needs for vehicle longevity than ever before. This brings to the fore a need to maintain peak performance over longer vehicle lifetimes, lest persisted inefficiencies drive total cost of ownership higher and increase environmental damage unnecessarily.\\ 

Core to the problem of persisted inefficiency is a lack of consumer knowledge. Though there have been efforts to make waste and emissions management more exciting~\cite{papamichael2022unified}, and therefore engaging, to average individuals, the same is less true for fields such as automotive maintenance. Without detailed knowledge of vehicles, or their maintenance needs, automobiles are often left to languish. There is an opportunity to build smarter vehicle diagnostics that help un-trained individuals assess, evaluate, and plan response to vehicle operating conditions.\\ 

While vehicle technology is rapidly advancing, fault diagnostics are lesser-explored, whether for trained individuals \textendash such as mechanics, or unskilled individuals \textendash such as vehicle owners and operators. There is a largely-unmet need for a better way to understand the state of vehicles such that operators might better access the information necessary to identify and plan response to impending or latent issues. At the same time, prior work shows that bringing expert knowledge to non-experts can have significant implications for fuel and energy savings~\cite{siegel2015smartphone,siegel_air_filter,siegel_misfire,siegel_tire_pressure} as well as safety~\cite{siegel_tire_cracks}.\\

Sound is an efficient, easy-to-use, and cost-effective means of capturing informative mechanical data~\cite{siegel2021surveying}, e.g. by using audio captured with a smartphone to identify fault or wear states~\cite{siegel_air_filter, siegel_misfire}. Compared with images~\cite{siegel_tire_cracks}, video, and other high-volume information, sound, like acceleration~\cite{siegel_tire_pressure,siegel2015smartphone}, is densely informative and compact. As a result, sound can be efficiently processed in near real-time by low-cost hardware such as embedded devices~\cite{siegel_afci}. Audio is particularly useful for providing insight into systems with periodic acoustic emissions, such as the rotating assemblies commonly found in vehicles and other heavy and industrial equipment. These data enable the identification and characterization of system attributes as well as fault diagnosis and preventive maintenance. This problem, however, is non-trivial and presents unique engineering challenges.\\ 

\begin{figure}[H]
\begin{center}
 \caption{Example of two acoustic vehicle samples as raw waveforms: inline four turbocharged (left) and a V8 with normal aspiration (right). Using the waveform to differentiate between vehicle types is a challenging task.}
 \includegraphics[width=0.45\textwidth]{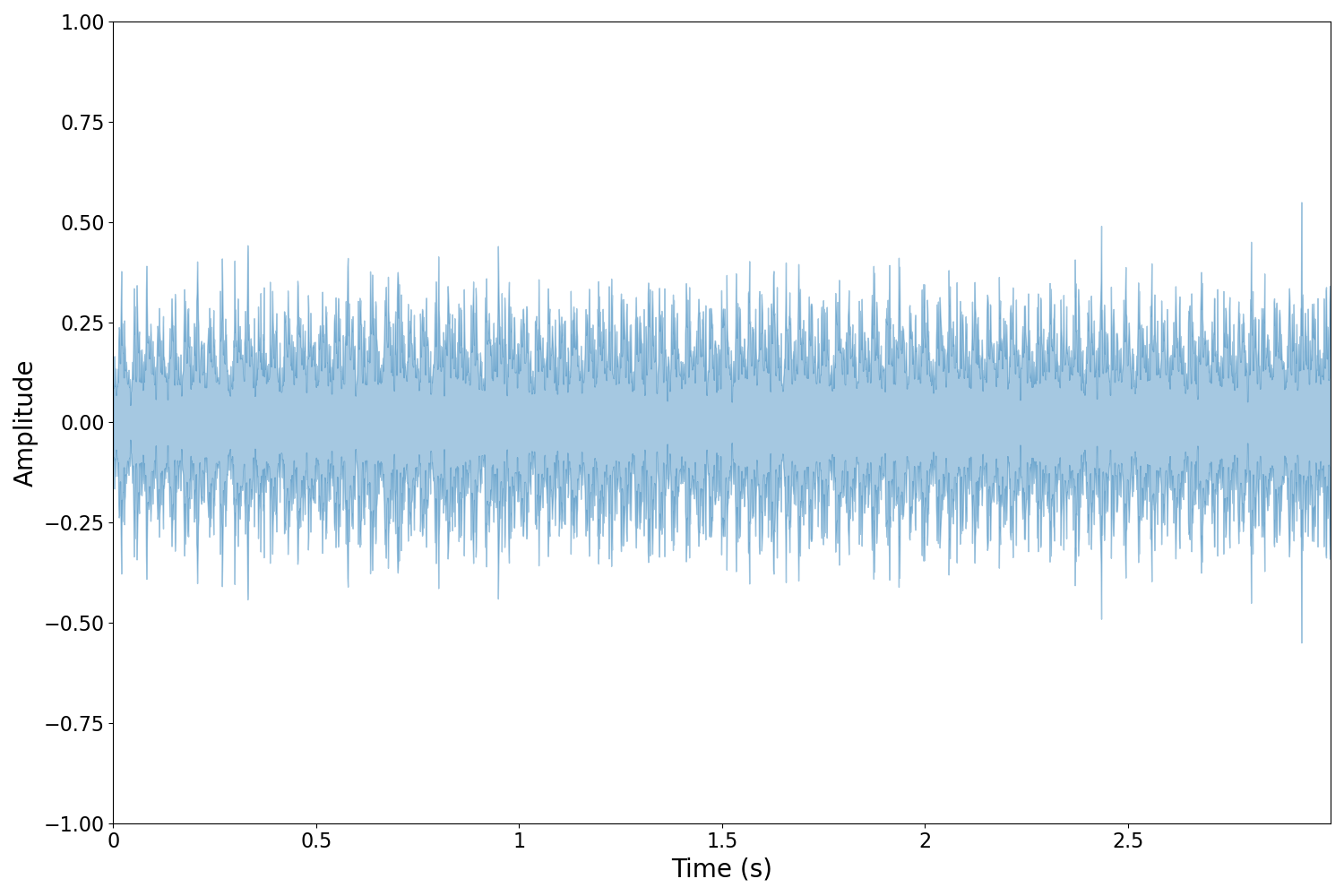}
 \includegraphics[width=0.45\textwidth]{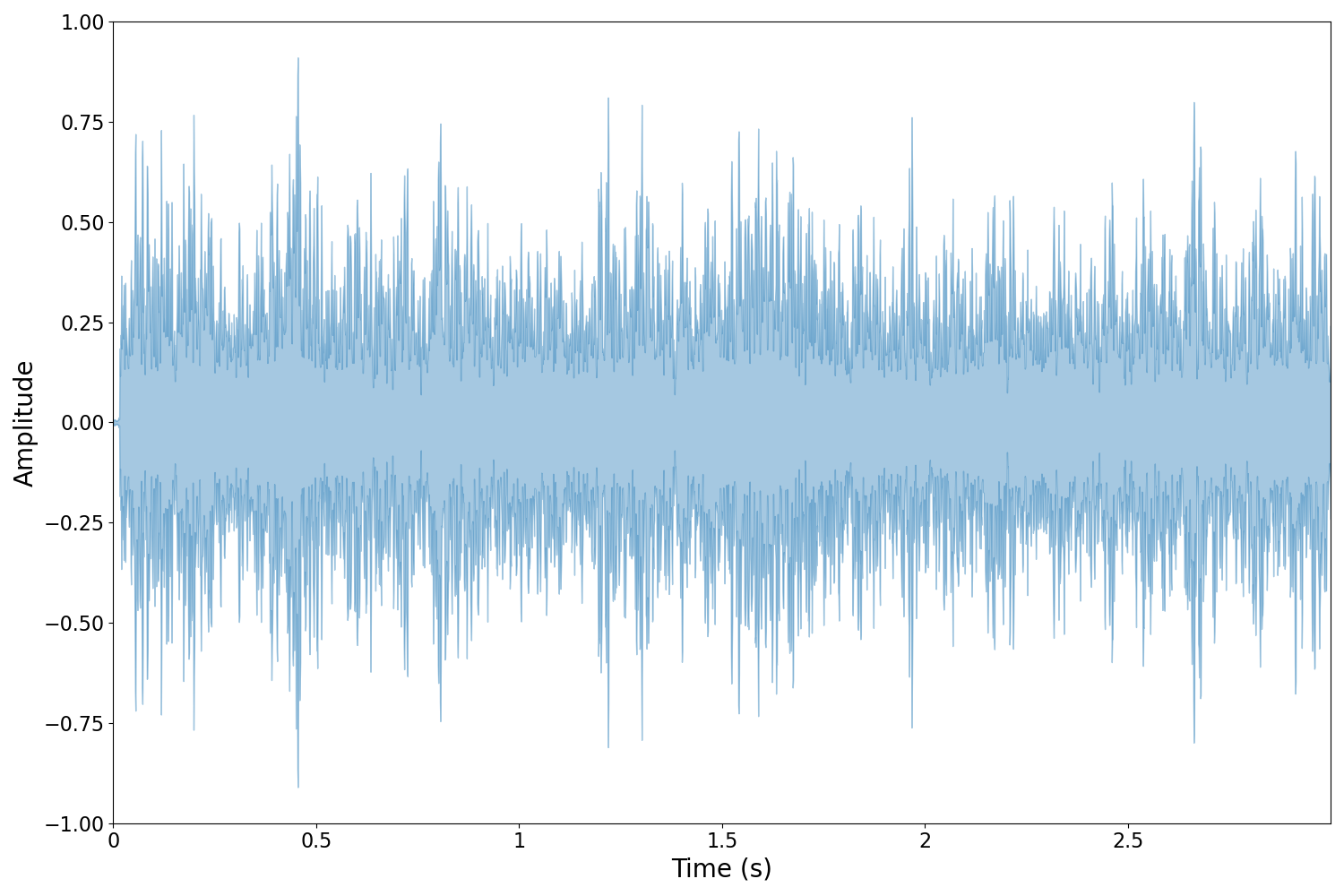}
 \label{fig:example-waveform}
 \end{center}
\end{figure}

\begin{figure}[H]
\begin{center}
 \caption{Example of two acoustic vehicle samples as spectrograms: inline four turbocharged (left) and a V8 with normal aspiration (right). Compared to the raw waveform from  Figure~\ref{fig:example-waveform}, the spectrogram provides a more informative feature type for vehicle differentiation.}
 \includegraphics[width=0.45\textwidth]{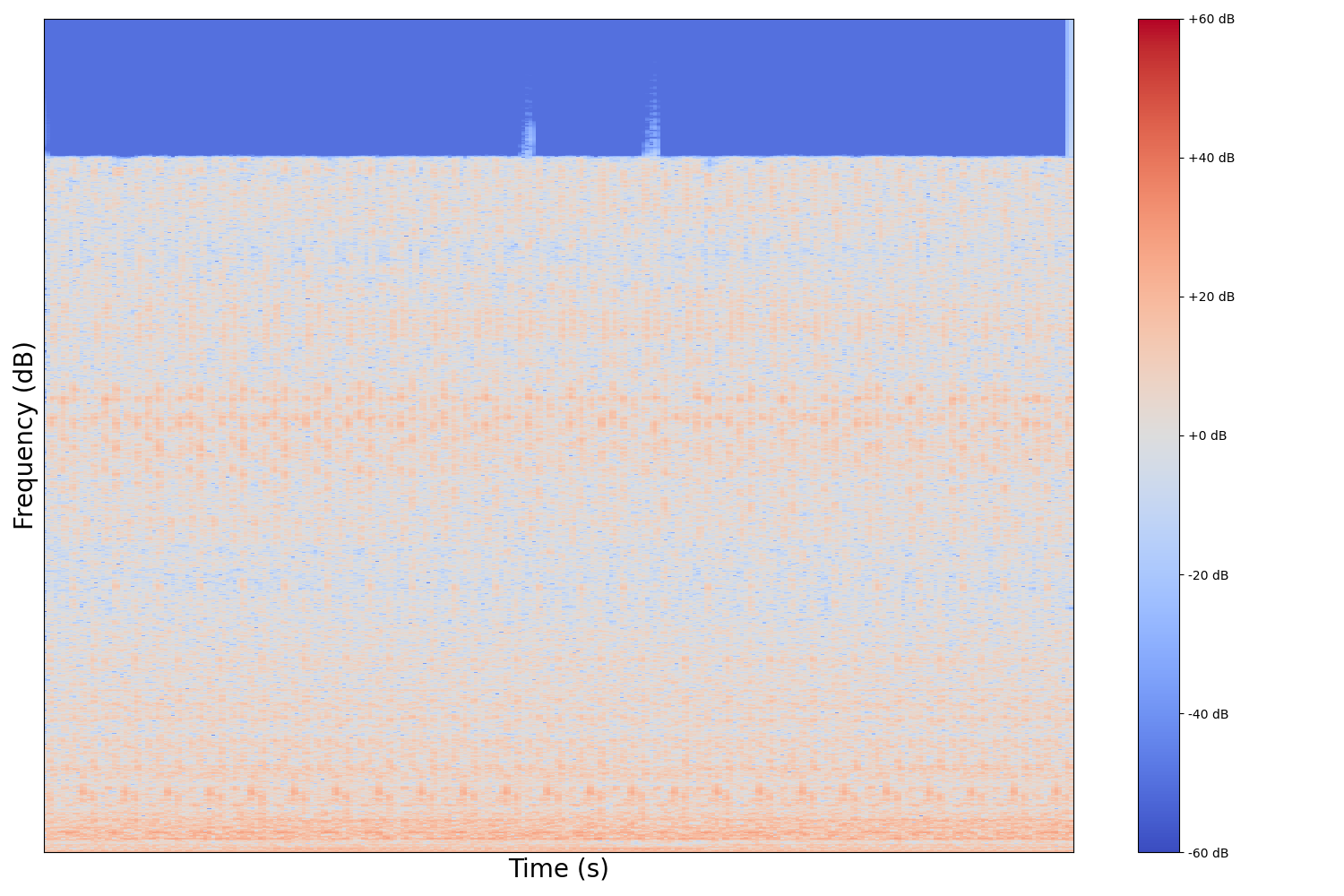}
 \includegraphics[width=0.45\textwidth]{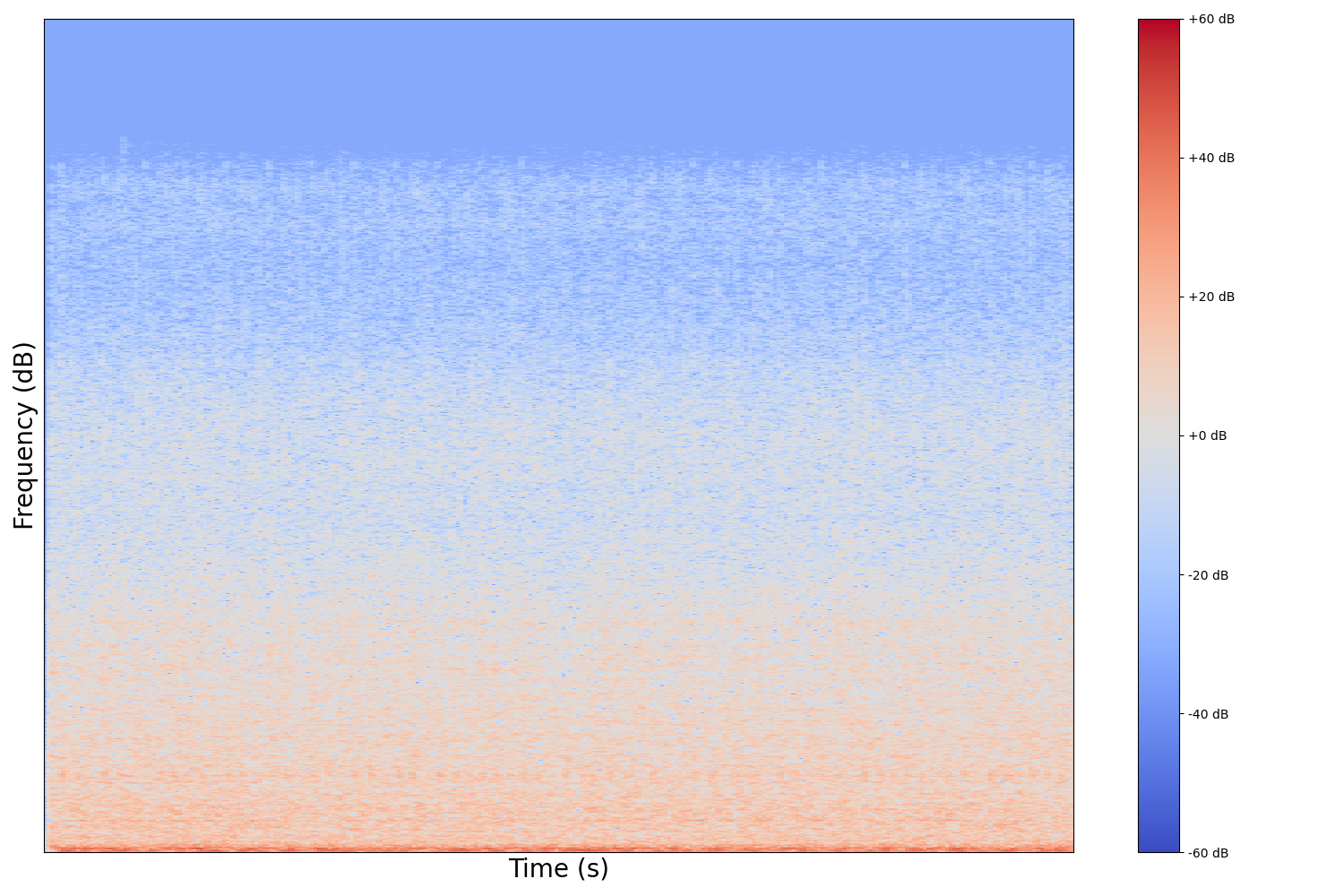}
 \label{fig:example-spec}
 \end{center}
\end{figure}

When characterizing an audio signal, the first feature explored is the raw sample itself, known as a waveform. This waveform provides standalone insights, though research has shown that feature extraction and transformation is a crucial step in building successful models in some acoustic recognition tasks~\cite{geiger2013large, sharma2020trends}. Transforming the raw waveform from the time domain to the frequency domain with the Fast Fourier Transform (FFT) provides particularly informative features. Other informative feature types utilize hybrid time and frequency information, such as Mel-Frequency Cepstral Coefficients (MFCCs), spectrograms, and wavelets. Figures \ref{fig:example-waveform} and \ref{fig:example-spec} show a representative waveform and spectrogram for two automotive engine types; note how the spectrogram is more visually-discernable. This characteristic similarly makes algorithmic differentiation an easier task.\\ 

Precise acoustic characterization algorithms benefit from prior knowledge of a sample's general class. In the context of vehicle diagnostics and characterization, identifying an audio sample as a vehicle is an initial step in classification. For example, ~\cite{audio_set} differentiates among water, road, rail, and air vehicles. Within road vehicles, as this work studies, attributes may include fuel type (gasoline or Diesel), engine configuration (flat, inline, V), cylinder count (4, 6, 8), or engine aspiration. Additional attributes of interest may include engine state (accelerating, idling, starting), make/model/OEM, horsepower, and so on. As proof of concept, in this work we focus on four main attributes: fuel type, engine configuration, cylinder count, and aspiration type. \\

A final challenge in acoustic vehicle understanding is status. In other words, we might ask whether a vehicle is in motion or stationary, or whether a vehicle is performing normally or abnormally. Given that it can take mechanics hours of manual labor and visual inspection to detect complicated issues with vehicles, it is clear how challenging \textendash and subjective \textendash this task may be using only sound data. In this work, we seek to identify whether a vehicle engine is operating normally or misfiring. \\ 

\begin{figure}[H]
\begin{center}
 \caption{Flowchart of our proposed cascading architecture for vehicle understanding using audio data. We can see there are four sequential levels with a conditional dependence between levels $1\rightarrow2$ and $3\rightarrow4$. Of particular note is the transition between level 2 and 3, as level 2 cascades its attribute predictions to level 3. This work focuses on level 2 and 3 as a proof-of-concept.}
 \includegraphics[width=0.60\textwidth]{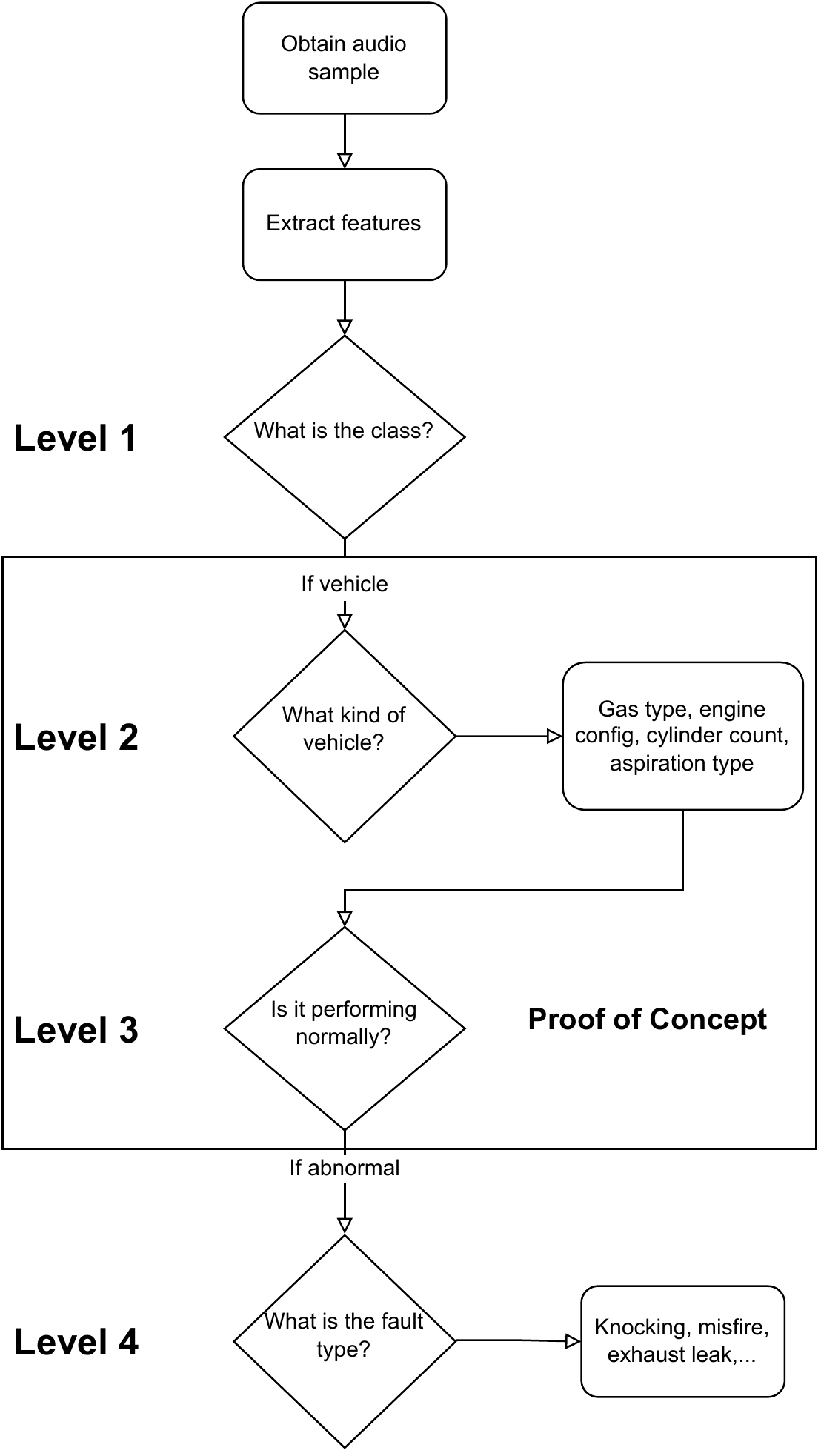}
 \label{fig:cascade_flowchart}
 \end{center}
\end{figure}

To address these challenges and build on the hypothesis that prior knowledge of a system's configuration can inform appropriate fine-grain classification, we propose novel cascading architectures for vehicle understanding, as visualized in  Figure~\ref{fig:cascade_flowchart}. We define a cascading architecture as a multi-level, sequential, conditional network that makes multiple predictions and cascades each prediction to every successive layer of the network. We also compare this architecture to a more conventional multi-task classification approaches. Our representative cascading network has four distinct layers:
\begin{enumerate}
 \item \textbf{General acoustic classification:} Does the audio sample contain a vehicle?
 \item \textbf{Attribute recognition:} What is the kind of vehicle?
 \item \textbf{Status prediction:} Is the vehicle performing normally?
 \item \textbf{Fault identification:} If abnormal, what fault is occurring?
\end{enumerate}

Our architecture is novel in that it integrates multiple highly-granular classification tasks and also in that the result of each successive classification task might inform the next. Previous work (surveyed in Section~\ref{sec:acoustic-vehicle-char}) has shown mastery of both high-level attribute recognition and low-level status prediction, though this article, to the best of our knowledge, is the first to complete these tasks simultaneously in a unified deep neural network network architecture.\\ 

In this manuscript, we apply audio data to vehicles as a means of providing enhanced insight towards the goal of reducing vehicle emissions and increasing usable service life. We first motivate the need and value for an automated system of vehicle understanding using sound data and then explore the prior art in sound recognition with deep learning, cascading architectures, acoustic vehicle characterization, and audio data augmentation. We define cascading architectures for vehicle understanding at a high level as multi-level, sequential, conditional networks as visualized in  Figure~\ref{fig:cascade_flowchart}.\\

In a proof-of-concept, we built a two-stage convolutional neural network (CNN) with its first stage specializing in vehicle attributes and then cascading its attribute predictions to the second stage misfire fault detection network. Through this approach, the cascading model achieves $95.6\%$, $93.7\%$, $94.0\%$, and $92.8\%$ validation accuracy on attributes (fuel type, engine configuration, cylinder count, aspiration type, respectively). The cascading CNN also achieved $93.6\%$ misfire fault validation accuracy which was $16.4\%$ and $4.2\%$ better than na{\"i}ve and parallel baselines, defined as predicting the most probable label and a one-stage multi-task CNN. Although the cascading model does not outperform the parallel baseline across all feature types in outsample test performance, we believe this nascent architecture has the potential to generalize well in real-world vehicle and other system classification scenarios with further refinement, optimization, and access to richer training data.\\

We provide evidence for the consistency and richness of our model architectures through experimental comparisons across three dimensions: models, features, and tasks. Additionally, we explore ablation studies focused on data augmentation, learning rate, and performance on outsample data captured from YouTube. We conclude with broader implications, future directions, and potential applications for this work.

\section{Prior art}
The presented work is novel in several areas:
\begin{enumerate}
 \item[2.1] Sound recognition with deep learning
 \item[2.2] Acoustic vehicle characterization
 \item[2.3] Cascading architectures
 \item[2.4] Audio data augmentation 
\end{enumerate}

To establish how the present work relates to and builds upon these areas, we now explore prior art in each.\\ 

\subsection{Sound recognition with deep learning} \label{sec:related-general}
There is breadth and depth in the history of deep learning~\cite{lecun2015deep, goodfellow2016deep} for sound-based tasks. One notable and highly recognizable sound-based task is automatic speech recognition (ASR), which looks to transcribe spoken language into written text. Traditional neural networks were first used for ASR~\cite{landauer1987learning, zavaliagkos1994speech}, later evolving to convolutional neural networks (CNNs)~\cite{abdel2014convolutional, palaz2015analysis} and recurrent neural networks (RNNs)~\cite{graves2013speech, graves2014towards}, with modern techniques gravitating towards transformers \cite{jaderberg2015spatial, dong2018speech, gulati2020conformer, wang2020transformer}. These differing neural network types have been utilized for diverse and unique sound-based tasks. \\

One such task that commonly leverages deep learning is general acoustic classification~\cite{hershey2017cnn}. We define this task as classifying a signal across many labels within broad categories. AudioSet~\cite{audio_set}, for example, is a popular dataset for general acoustic classification, which contains 527 diverse, hand-annotated classes for 2.1 million YouTube videos. AudioSet~\cite{audio_set} provides an ontology for general acoustic classification with seven categories including:
\begin{itemize}
 \item human sounds (speech, hands, bodily functions)
 \item animal (domestic, farm, wild)
 \item music (instrument, genre, mood)
 \item sounds of things (vehicle, engine, tools, alarms)
 \item natural (wind, water, storms)
 \item source-ambiguous
 \item channel, environment, background
\end{itemize}

Related prior art focuses on high-level audio classification using neural networks \cite{battaglino2016acoustic, bae2016acoustic, mcloughlin2017continuous, sharan2017robust, weiping2017acoustic, aslam2018acoustic, huang2019acoustic, abesser2020review}. These works often utilize audio embeddings~\cite{arandjelovic2017look, cramer2019look, kong2020panns, gong2021ast} with transfer learning to better inform the overall acoustic classification task. As observed with the Audioset~\cite{audio_set} ontology, there is also a diverse world of sound that extends to low-level fine-grained task prediction. Applications that utilize DL for sound-based tasks range from music genre classification ~\cite{jeong2016learning, oramas2018multimodal} to medical diagnoses~\cite{chen2016s1, dascalu2019skin, brown2020exploring} to animal recognition~\cite{mac2018bat, stowell2019automatic, xie2019handcrafted, zhong2020beluga, zgank2021iot, bravo2021bioacoustic} and more~\cite{liu2016rolling, torres2017baby, lee2020intelligent}.\\

We build on similar classification techniques and differentiate our work in two ways: first, we focus on multiple low-level, fine-grained multi-level label prediction with the assumption of the highest-level class, rather than simply classifying a single level at a time. Given our approach focuses on these fine-grained tasks, we do not utilize general acoustic embeddings~\cite{arandjelovic2017look, kong2020panns, gong2021ast} but rather train lightweight models from scratch.\\ 

Second, we utilize data collected at a higher sampling rate ($48$~kHz) compared to that collected from YouTube, which we show Section~\ref{sec:ablation-YT} caps its sampling rate and in the process may discard useful informative features. This is important because models trained on such crowd-sourced audio may not perform as well as models trained and operated using raw, full-frequency audio directly collected from a mobile device. YouTube and other such sources also conduct feature-destructive compression on some audio samples, which can limit model performance and generalizability.\\ 

In summary, our approach uses larger frequency ranges to provide detailed classification lower into the stack, building on techniques demonstrated by prior art.

\subsection{Acoustic Vehicle Characterization} \label{sec:acoustic-vehicle-char}
Our work is novel in its application of sound-based deep learning, as we focus on acoustic vehicle characterization. Prior art \cite{siegel2021surveying} has conducted an extensive survey on off-board vehicle vibro-acoustic diagnostics which explores a multitude of acceleration- and sound-based vehicle characterization approaches. Our work focuses on the lesser-explored tasks of vehicle attribute recognition, engine misfire fault detection, and specifically the interrelation between vehicle variant and fault-specific diagnostic algorithms.\\

Previous work has focused on vehicle attribute recognition~\cite{umberto_thesis, siegel2021surveying} utilizing traditional ML techniques such as SVMs and feature selection. These primarily consider the FFT and MFCC feature types, whereas our work considers five distinct feature types and conducts an in-depth feature comparison. These works also do not consider the engine configuration attribute task. Our work explores the three previously explored attribute prediction tasks: fuel type, cylinder count, aspiration type, as well as the engine configuration task. \cite{umberto_thesis, siegel2021surveying} also build separate models for each prediction task, whereas our deep learning approach demonstrates that predicting attributes jointly in a single model can yield strong results, with variant-identifying parameters helping to ``shape'' the final diagnostic model.\\ 

One research study looking to identify specific faults~\cite{siegel_misfire} collects nearly $1k$ audio samples of vehicles using smartphone microphones for which approximately $1/3$ are abnormal ``vehicle misfire'' events. Using traditional machine learning techniques and various audio features combined with feature selection, \cite{siegel_misfire} achieves $1.0\%$ misclassification rate and $1.6\%$ false positive rate. However, this previous work was trained on three vehicles types and it is unclear if and how the results would generalize to a broader sample including more diverse vehicles.\\

Our work is unique in its merging of vehicle acoustic characterization tasks. First, classifying attributes for an abnormally performing vehicle may be a more challenging task and \cite{umberto_thesis, siegel2021surveying} only focused on normally performing vehicles. Second, we show the value of multi-task learning through attributes prediction improving misfire performance and vice versa. Finally, we demonstrate a proof-of-concept for a cascading architecture by with our two-stage CNN which uses the first stage to specialize on attributes and then cascade its attribute predictions to the second stage misfire detection network.

\subsection{Cascading architectures} \label{related:cascade}
We envision our cascading architecture for vehicle understanding as a sequential, conditional model. Specifically, we construct an architecture with four sequential levels (Figure~\ref{fig:cascade_flowchart}), where levels $1\rightarrow2$ and $3\rightarrow4$ are conditionally dependent. Additionally, level 3 is dependent upon the cascaded result from level 2.\\

There is valuable related work in sequential and conditional deep learning that has been utilized in a select applications. Examples include human activity recognition~\cite{yang2017scnn}, image segmentation~\cite{wang2019learning}, facial recognition~\cite{Xiong_2015_ICCV}, frame prediction in games~\cite{wang2017deep}, and music rhythm generation~\cite{makris2018deepdrum}.

There are three works of particular relevance to our cascading architecture for acoustic vehicle characterization:
\begin{itemize}
 \item An Ontology-Aware Framework for Audio Event Classification~\cite{sun2020ontology}
 \item Masked Conditional Neural Networks for sound classification~\cite{medhat2020masked}
 \item CascadeML: An Automatic Neural Network Architecture Evolution and Training Algorithm for Multi-label Classification~\cite{pakrashi2019cascademl}
\end{itemize}

We now discuss how our work differs from these approaches.\\

\cite{sun2020ontology} has elements that are similar to our proposed cascading architecture. Specifically, \cite{sun2020ontology} leverages the relationship between fine and coarse labels in their model training. The first and second level of our cascading architecture can be seen as examples of coarse and fine labels, which have a subset relationship. However, \cite{sun2020ontology} does not consider conditional logic and does not have multiple levels like our proposed cascading architecture.\\

Both \cite{medhat2020masked} and our work utilize neural networks for acoustic classification. However, the conditional component of \cite{medhat2020masked} does not correspond with multi-label dependence proposed in our cascading architecture but rather conditional dependence in the data itself. Specifically, \cite{medhat2020masked} leverages the temporal relationship of nearby frames in spectrograms whereas our work utilizes the entire spectrogram, agnostic to its temporal nature, as a feature set in the CNN.\\

CascadeML \cite{pakrashi2019cascademl} is closely related to our proposed architecture. Our architecture might almost be considered an example of the theoretic structure proposed by \cite{pakrashi2019cascademl}. However, \cite{pakrashi2019cascademl} focuses on the AutoML component where the network can dynamically grow itself based upon any number of multi-label samples. We instead embed expert knowledge into the static structure of the cascading architecture. In other words, we know fault type is conditional upon status prediction, so we embed that in the structure of our architecture rather than building a network from scratch with the AutoML approach of \cite{pakrashi2019cascademl}. Not only will this approach allow us to demonstrate stronger and more consistent performance, but also it would be theoretically more efficient given it structure is already defined.\\

In summary, \cite{sun2020ontology,medhat2020masked,pakrashi2019cascademl} showcase various sequential, conditional, and cascading architectures that are closely related to our work. However, these works focus on either theoretical or high level acoustic classification. We show there is value in building a model for the specific application of acoustic vehicle characterization.

\subsection{Audio Data Augmentation} \label{sec:related-data-augmentation}
In deep learning for computer vision, there is value applying data augmentation to images via rotations, cropping, scaling, etc. to gain invariance among unique inputs. Audio may similarly be augmented.  \cite{salamon2017deep} utilized time stretching, dynamic range compression, pitch shifting, and background noise for environmental sound classification. In speech recognition, \cite{ko2015audio} applied multiple speed changes to expand their training set. Focused on animal audio classification, \cite{nanni2020data} uniquely applies computer vision augmentation techniques to the spectrogram such as reflection and rotation, as well as waveform augmentation with pitch, speed, volume, random noise, and time shift. For music classification, \cite{schluter2015exploring} found pitch-shifting to be most informative but also utilizes time stretching and random frequency filtering.\\

There is prior art involving data augmentation for acoustic vehicle characterization. Researchers \cite{yang2019audio} classify accelerating vehicles using audio and apply random noise and pitch shift to expand their dataset. Other work \cite{chen2021hybrid} focuses on vehicle type identification using MFCCs and other audio features, these authors apply city noise to augment captured signals. There have been additional applications of augmentation to acoustic vehicle DL related tasks \cite{barak2019audio, bu2021adversarial, sharif2021imobilakou}.\\

Our work is similar to \cite{schluter2015exploring, salamon2017deep, nanni2020data} in that we utilize pitch, speed, volume, and random noise augmentation types. However, we differentiate ourselves in our application of the augmentation to multi-task learning for vehicle attribute recognition and fault detection. Additionally, to our knowledge, our approach is unique in that our train set is not a combination of real and augmented samples but is entirely made up of augmented samples. We hypothesize this may allow for more robust training and less overfitting / memorizing of specific samples. We also show in Section~\ref{sec:ablation-aug} the strong, positive impact augmentation has on our model performance.

\section{Approach}

As noted in the problem motivation, proper maintenance is increasingly important to vehicles as their number and miles travelled grow. However, as vehicles become more like appliances and less familiar to their users, driver and passenger awareness of their capabilities diminishes. We suggest that automated vehicle identification is an important first step in diagnosing and rectifying problems plaguing efficiency and emissions: when a part fails, it is not enough to know that a vehicle is big or small, or red or silver, but rather that a vehicle is a 2016 version of some specific make and model possessing a particular engine and transmission type and tire size. Creating tools that effectively identify vehicles and use these attributes as a means of improving state characterization allows non-expert users to expertly diagnose their own vehicles, clearing the knowledge hurdle to effective maintenance and thereby reducing fuel consumption, emissions, and likelihood of breakdown.\\

Our approach is a proof-of-concept for how vehicle characteristics can better inform fault diagnoses and in this section we explain our approach from five main angles:
\begin{enumerate}
 \item[3.1] Datasets
 \item[3.2] Features
 \item[3.3] Model
 \item[3.4] Network
 \item[3.5] Implementation 
\end{enumerate}
First, we share the details of our novel dataset, specifying the dataset splits and augmentation procedure. Second, we compare the different audio feature types used in our models. Third, we develop a high-level understanding of the two multi-task models proposed in this work. Then, we clarify the specifics of the CNNs for each main feature type. Finally, we provide implementation details to aid in reproducibility. 

\subsection{Datasets}
Our dataset is a collection of audio samples recorded using mobile phones. We utilize mobile phone microphones as these tend to have repeatable characteristics within the range of frequency similar to those of human speech and hearing (roughly $20$Hz-$20$kHz). In particular, we merge the datasets from~\cite{umberto_thesis, siegel2021surveying}, which focus on attributes, with \cite{siegel_misfire, siegel_harvard_dataset}, which focus on misfire detection. We outline the makeup of our dataset and its splits in Table~\ref{tab:datasets}.\\

\begin{table}[H]
\caption{Summary of the train/validation/test split for our datasets. We create 8 augmented training samples per one validation sample. The test sets is mutually exclusive from the train and validation sets, which allows it to provide an insight into the outsample performance and model generalizability.}
\label{tab:datasets}
\begin{center}
\begin{tabular}{|c|c|c|c|} \hline
\textbf{Dataset Split} & \textbf{Raw Samples} & \textbf{Clips (3 sec)} & \textbf{Total Length (min)}\\ \hline
Train & -- & 43640 & 2182.0 \\
Validation & 228 & 5455 & 272.75\\
Test & 58 & 951 & 47.55 \\
\midrule
\textbf{Total} & \textbf{286} & \textbf{50046} & \textbf{2502.3} \\
\hline
\end{tabular}
\end{center}
\end{table}

Together, these datasets totaled $286$ ($228+58$) audio samples, which contributed to about 320 minutes of non-augmented audio. We decided to split each audio sample into three second clips (``chunks'') to provide our features and models with a standard input size for training and to expand our dataset size. This time was selected as striking an appropriate balance of capturing low-frequency sounds with ease and speed of capture. Additionally, we believe three seconds is a long enough sample to be representative data for our model, but also short enough that would allow for clips within the same sample to be distinct and useful for training.  After chunking the samples, this resulted in $6406$ ($5455+951$) unique three second clips. Though fixed-length clips were used, our models at test time are agnostic to the input feature size, rather than needing the sample to be exactly three seconds.\\

Our approach for the train/validation/test split involved building a train set using only augmented clips with the goal of reducing overfitting and improving generalizability. In turn, we split the real full length samples into a validation and test set, using a traditional 80/20 split. As seen in Table \ref{tab:datasets}, this resulted in $5455$ validation clips and $951$ test clips. We split the validation and test set using the raw samples rather then the clips because we created the train set using only the validation set. This allowed for the test set to be mutually exclusive from both the train and validation set to showcase as accurate as possible the model's outsample generalizability.\\

\begin{figure}[H]
\begin{center}
 \caption{Label distribution for the validation set. We observe the classes are not perfectly balanced and some labels in the engine configuration and cylinder count are very underrepresented.}
 \includegraphics[width=1.0\textwidth]{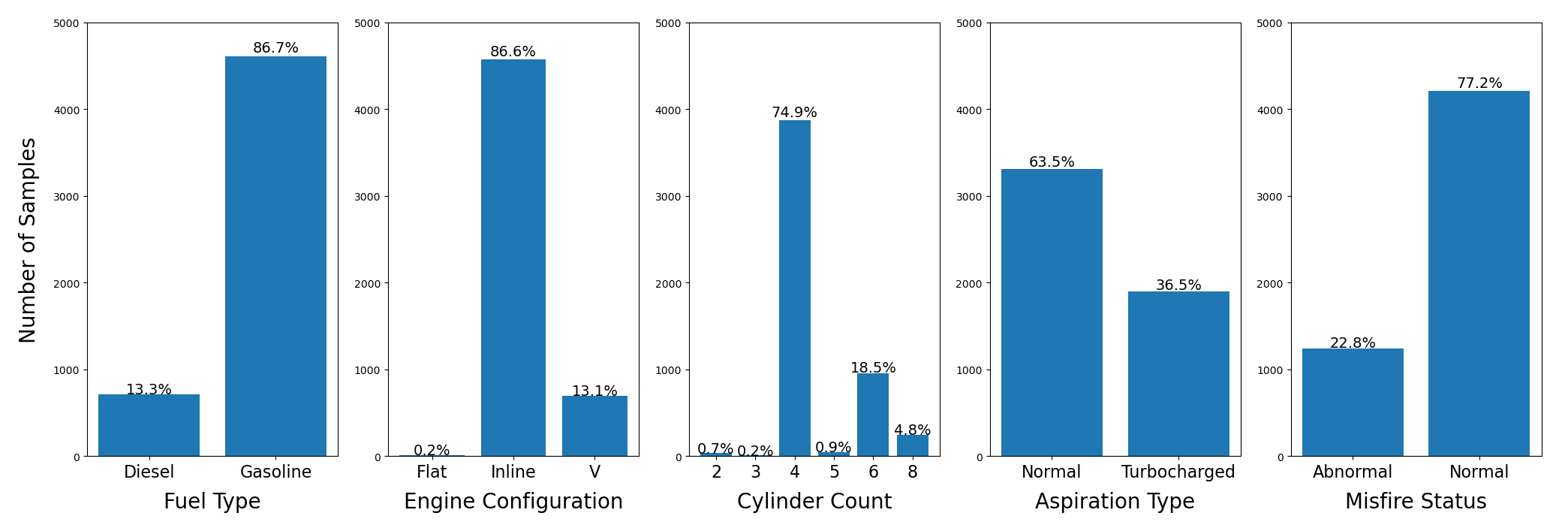}
 \label{fig:label-val}
 \end{center}
\end{figure}

\begin{figure}[H]
\begin{center}
 \caption{Label distribution for the test set. We attempted to build a test set with as similar label distribution to training and validation as possible, while also including at least one raw sample from each label.}
 \includegraphics[width=1.0\textwidth]{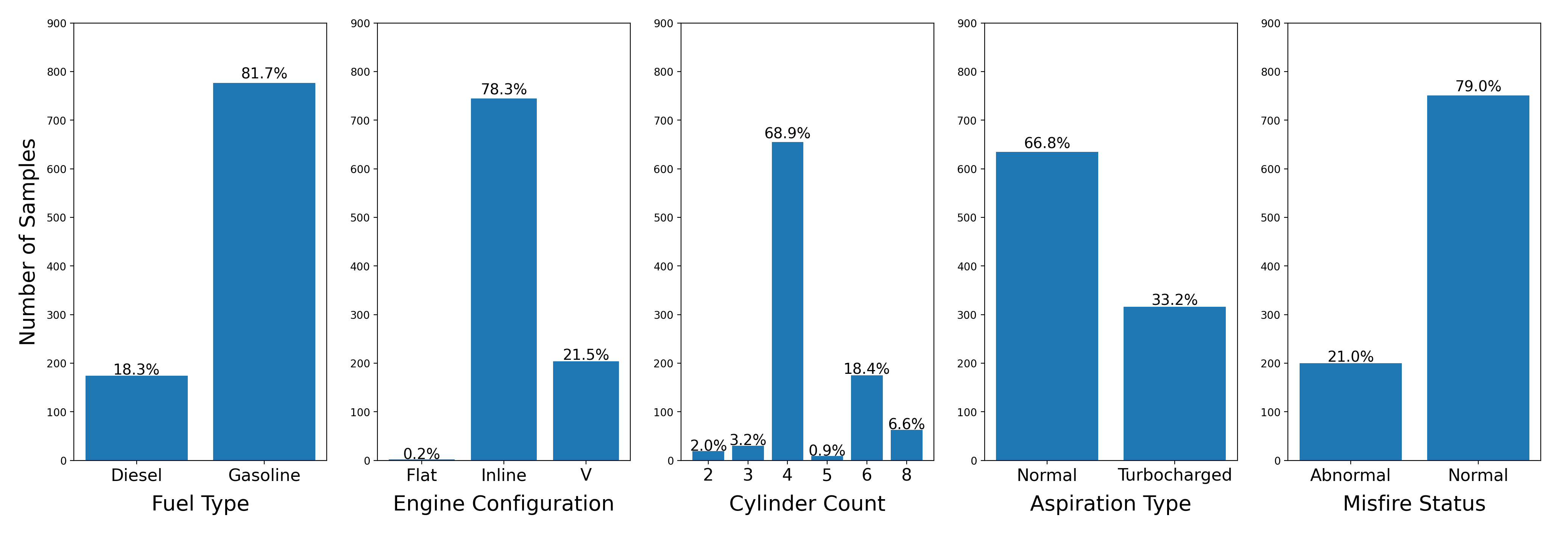}
 \label{fig:label-test}
 \end{center}
\end{figure}

Due to the relatively small size of the datasets, we manually created the validation and test splits with an attempt to balance the label distribution to the best of our ability for the four attribute recognition tasks: fuel type, engine configuration, cylinder count, and aspiration type and the status prediction task focused on misfire detection. Label distributions are visualized in Figures \ref{fig:label-val} and \ref{fig:label-test}.

Our process of creating the training set consisted of applying a data augmentation process eight times to each of the $5455$ validation clips, which when expanded resulted in a train set of size $43640$ as seen in Table \ref{tab:datasets}. The data augmentation process consisted of applying four types of data augmentation to each clips which each data augmentation type having a probability of $50\%$ of being activated. The types of augmentation chosen were volume change, pitch shift, speed shift, and random background noise. More specifically, we chose each type of augmentation from a randomly uniformly distribution over these augmentation parameters:
\begin{itemize}
 \item Change volume between -5.0 and 5.0
 \item Pitch shift between -0.25 and 0.25
 \item Change speed between 0.92 and 1.08
 \item Add background noise with signal to noise ratio (SNR) between 0.05 and 0.20.
\end{itemize}

We also add epsilon between $10^{-6}$ and $10^{-5}$ to each parameter, so as to not randomly choose a hyperparameter of zero volume change / pitch shift or one for speed change, which would result in the clip not being augmented at all. Additionally, if the clip was not changed after all four augmentation types rolled their probability (i.e. all four types were not activated), we applied one augmentation type at random. This guaranteed that no training clips were exactly the same as their respective validation clip. All value ranges were determined experimentally, with expert mechanic review to coarsely validate the ``closeness'' of the augmented sample with the original sample. Augmentation parameters were set so as to line up with typical variation in engine configurations, e.g. through manufacturing diversity and wear. For example, frequency shift was bounded based on typical allowable tolerances for idle speed. Amplitude limits were set so as to minimize the effect of signal clipping. The features were determined to be acoustically-discernible after augmentation.\\

We further motivate our decision for using data augmentation later in Section~\ref{sec:ablation-aug}. \\

In summary, we created a new dataset with a non-traditional train/validation/test split. This set uses only augmented variants of validation samples in training and is mutually exclusive from both the validation and test set. Our dataset, per Table \ref{tab:datasets}, in total consisted of $50046$ three second clips for a total of $2502.3$ minutes or about $42$ hours total of audio.

\subsection{Features}
In this work, we explore four audio feature types (five when considering the raw waveform) which include:
\begin{itemize}
 \item Fast Fourier Transform (FFT)
 \item Mel-frequency Cepstral Coefficients (MFCCs)
 \item Spectrograms
 \item Wavelets
\end{itemize}

\begin{figure}[H]
\begin{center}
 \caption{Example of FFT audio features, which are binned by frequency (24,000 bins which correspond to the 24kHz sampling rate and represented with a long 1D vector.}
 \includegraphics[width=0.80\textwidth]{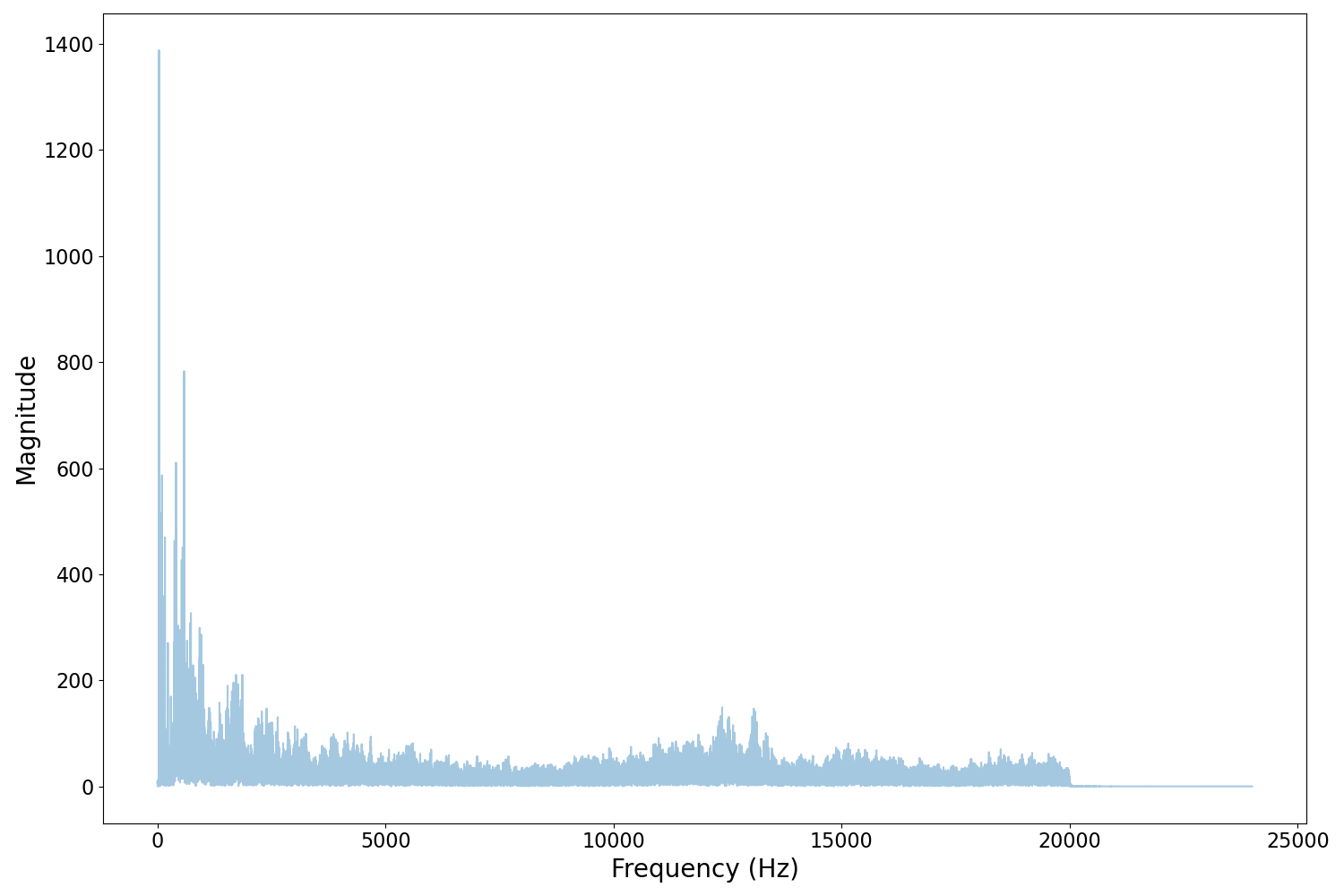}
 \label{fig:example-fft}
 \end{center}
\end{figure}

\begin{figure}[H]
\begin{center}
 \caption{Example of wavelet audio features, which use to 13 levels of decomposition and then are stacked horizontally to create one long 1D vector.}
 \includegraphics[width=0.60\textwidth]{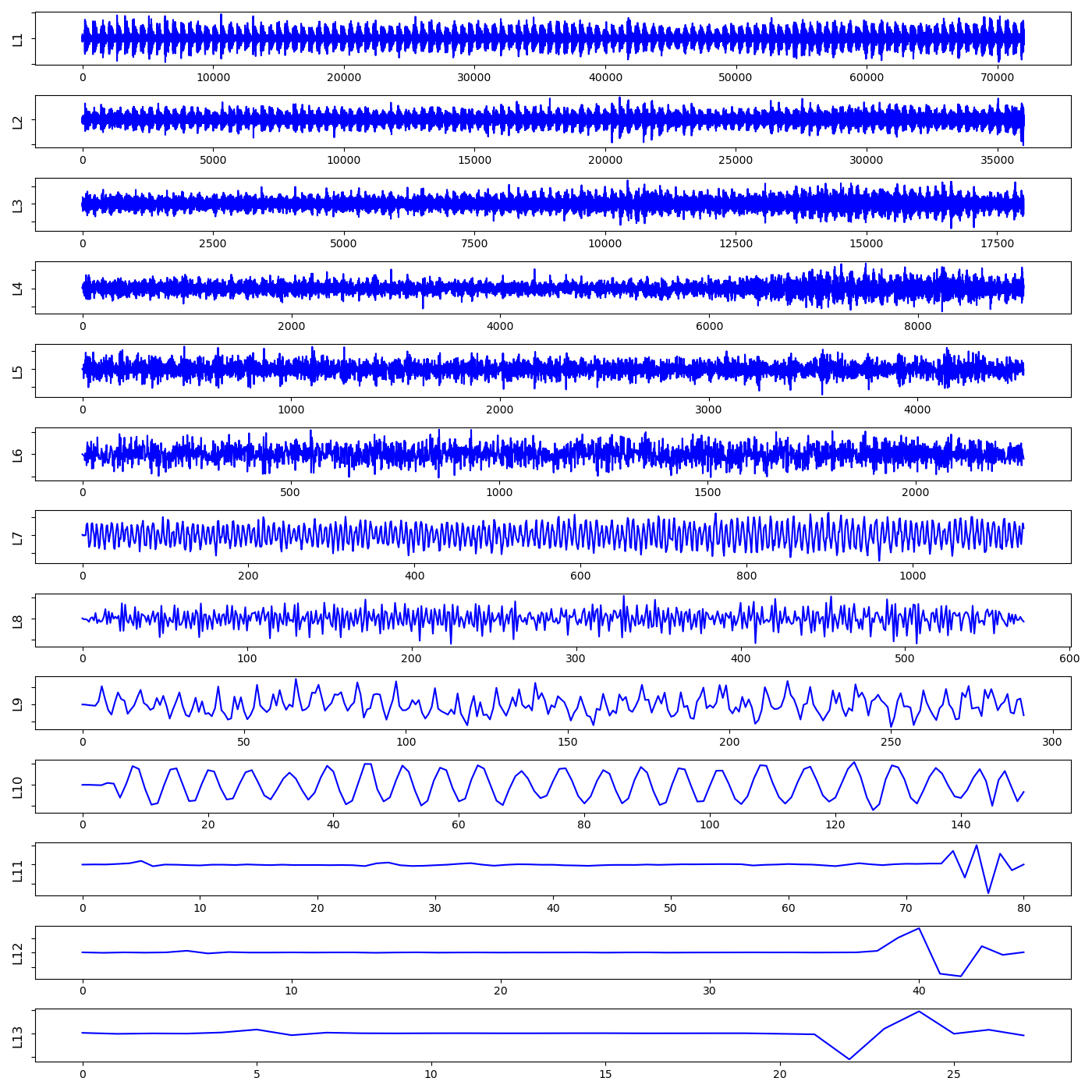}
 \label{fig:example-wavelets}
 \end{center}
\end{figure}

\begin{figure}[H]
\begin{center}
 \caption{Example of MFCC audio features for which 13 coefficients were chosen as a hyperparameter. MFCC features are later shown to be the more impactful in our deep learning models.}
  \includegraphics[width=0.80\textwidth]{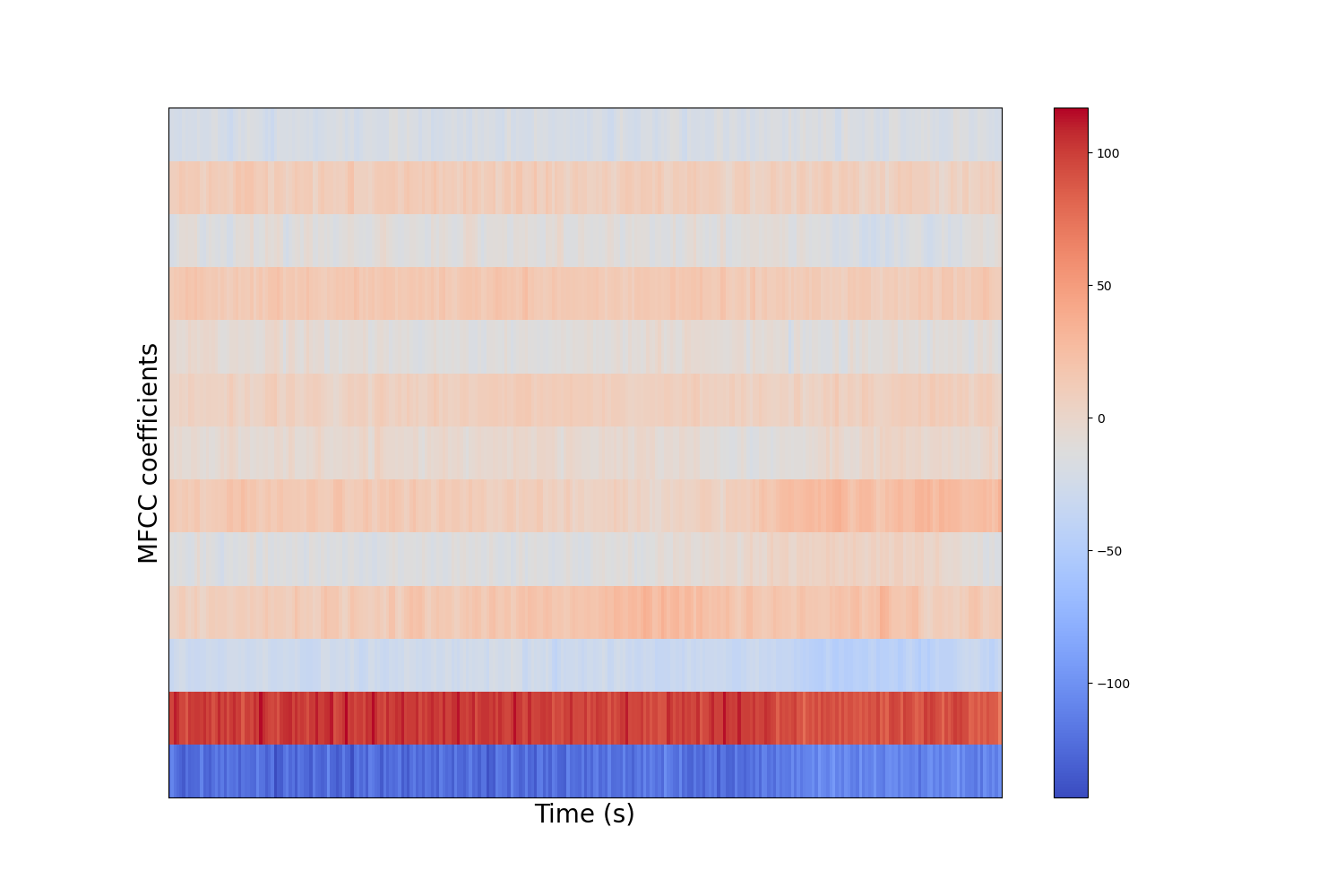}
 \label{fig:example-mfcc}
 \end{center}
\end{figure}

Examples of these feature types can be visualized in Figures \ref{fig:example-waveform}-\ref{fig:example-spec} and \ref{fig:example-fft}-\ref{fig:example-mfcc}.\\

These features were chosen to give our models a diverse set of inputs. The raw waveform provides the model with time information, while the FFT provides the model with frequency information. MFCCs, Spectrograms, and Wavelets provide the model with varying degrees of hybrid time and frequency information at different dimensionality.\\

\subsection{Model}
We design and implement a variant of the earlier-proposed Cascade model: a two-stage network for which the first stage specializes on attributes classification and the second stage specializes on misfire detection. The first stage receives one of the previously described feature sets as input. The second stage receives both the feature set and the attribute predictions from the first stage. Both stages are trained jointly using multi-task learning.\\

To compare against the Cascade model, we also consider two baselines: Na{\"i}ve and Parallel models. The Na{\"i}ve model is a straightforward baseline that predicts the most represented class for each task. For example, if we look at  Figure~\ref{fig:label-val}, the Na{\"i}ve model would always predict gasoline as the fuel type. This model would achieve $86.7\%$ validation accuracy.\\

\begin{figure}[H]
\begin{center}
 \caption{Flowchart of our baseline Parallel model architecture. The Parallel model is differentiated from the Cascade model in that rather than level 2 cascading to level 3, we combine these levels in parallel.}
 \includegraphics[width=0.75\textwidth]{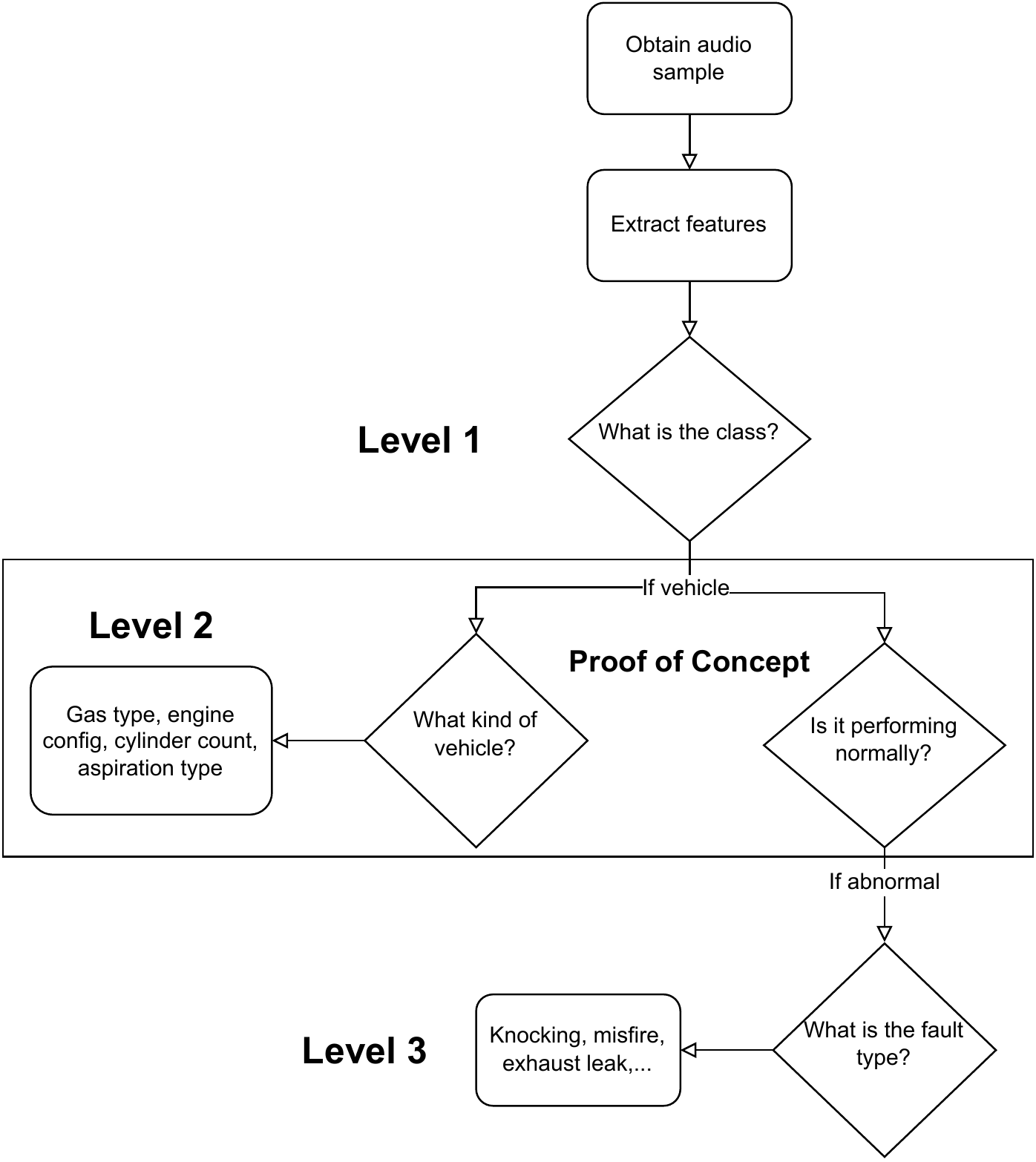}
 \label{fig:parallel_flowchart}
 \end{center}
\end{figure}

The Parallel model, as seen in Figure~\ref{fig:parallel_flowchart}, represents a more complex baseline to consider than the Na{\"i}ve model. The Parallel model is represented as a sequential, multi-level, conditional network in the same archetype as the Cascade model. However, per Figure~\ref{fig:parallel_flowchart}, it does not explicitly cascade, or provide as input, the classified vehicle type to the status prediction level. Rather, these two tasks are predicted in parallel from the raw waveform and/or features.\\

Both the Cascade and Parallel model are viable classification approaches: both serve as a proof-of-concept for a full scale AI mechanic architecture and provide an all-in-one solution for classifying attributes and misfire. The Cascade and Parallel models also both utilize multi-task learning to allow for the attributes prediction to better inform the misfire prediction and vice versa. However, the models differ in their internal sharing of intermediate classification representations. Therefore, there is value in the comparison on whether a joint, shared representation via the Parallel model or separate, specialized representations via the Cascade model would yield better performance. It is also worth noting that given the Cascade model has a second sequential stage, there is an non-negligible increase in model runtime and memory. In turn, these factors should be worth considering if the Cascade model provides a margin on any task predictions.\\

\begin{figure}[H]
\begin{center}
 \caption{Parallel and Cascade CNN model architecture comparison. We observe the Parallel model uses a shared representation for both the attributes and misfire predictions whereas the Cascade model has two distinct CNNs that allow for specialization in attributes and misfire. Additionally, both networks utilize multi-task learning so the attributes loss can inform the misfire loss and vice versa.}
 \includegraphics[width=0.75\textwidth]{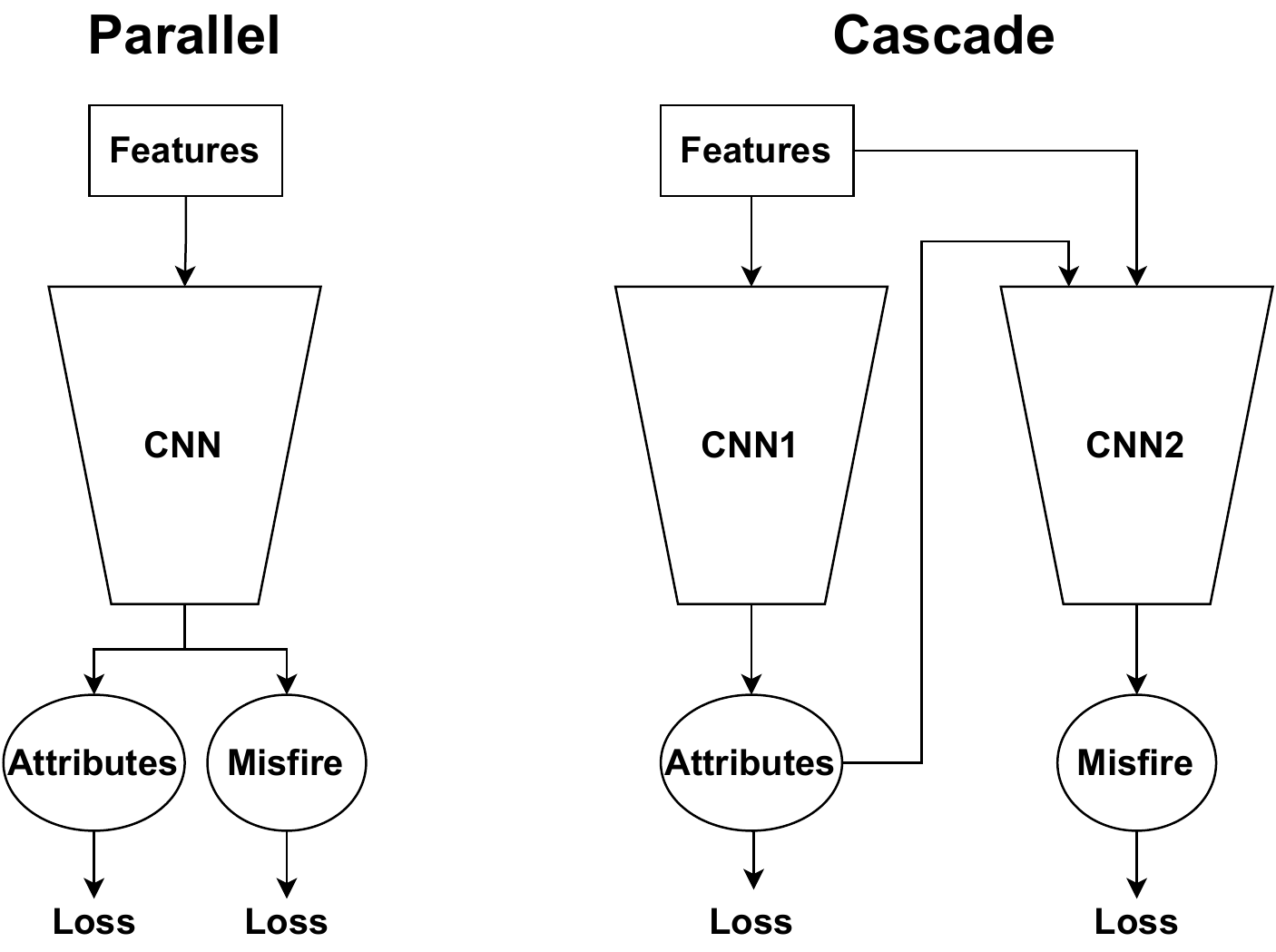}
 \label{fig:comparison}
 \end{center}
\end{figure}

These two models are directly compared in Figure~\ref{fig:comparison}.

\subsection{Network} \label{sec:network}
We propose two distinct multi-layer convolutional neural networks (CNNs), Parallel and Cascade, to predict both vehicle attributes and engine misfire, as seen in Figure~\ref{fig:comparison}.\\

For each, we group audio features into 1D (FFT, waveform, wavelets) and 2D (Spectrogram, MFCCs) sets, with two two distinct model architectures that utilize 1D and 2D convolution respectively. We adapt the M5 architecture~\cite{m5} which was built for general audio classification using the raw waveform.\\

Our samples utilize a $48$~kHz sampling rate, for which we consider frequencies $\leq 24$~kHz according to the Nyquist-Shannon Sampling Theorem. These samples were collected using a stereo microphone for which we average the dual channel input into a single mono channel. We split each sample into 3 second chunks which results in a $1\times72,000$ input vector for the raw waveform.\\

The unique aspect of the M5 model utilizes a large convolutional kernel size of $80$ in the first layer to propagate a large receptive field through the network. We adopt kernel size of $3$ from~\cite{m5} for the remaining three convolutional layers.\\

For the MFCC model we utilize kernel size of $2\times2$ since we chose $13$ MFCC coefficients as a hyperparameter which results in an input dimensionality of $130\times13$. For the spectrogram, we choose hyperparameters of $512$ for hop length and $2048$ for window size. This results in a input dimensionality of $1025\times282$. Since it was a larger input dimension than the MFCCs, we utilize traditional $3\times3$ kernels for spectrogram.\\

For the 2D models, our model consists of three convolution layers, rather than four layers for the 1D models because of their smaller input width. We follow the precedent set by~\cite{m5} for including pooling, batchnorm, and ReLU after each conv layer in both 1D and 2D models. We also add dropout with probability $= 0.5$ after each layer to improve generalizability and to minimize the likelihood of overfitting. We treat each prediction task as classification and therefore have final fully connected output layers with dimensionality corresponding to the number of classes for each task. These predictions are then fed into log softmax and trained using negative log-likelihood loss.

\subsection{Implementation}
We built our models using PyTorch within an Anaconda environment. We trained our models using 1080 Ti and Titan XP GPUs. Our models unless otherwise stated were trained for $100$ epochs with the Adam optimizer, utilizing a batch size of $128$, learning rate of $0.001$, no weight decay, and with amsgrad following the standard set by \cite{m5} for training CNNs on audio. The Spectrogram model did not fit into GPU memory with a batch size of $128$, so a batch size of $64$ and $32$ were used for the Parallel and Cascade models, respectively.\\

We conducted a handful of hyperparameter searches to find the best set for each model type, across number of epochs, learning rate, and dropout rate. Specifically, we looked at number of epochs ranging from $10$ to $100$, learning rate ranging from $0.0001$ to $0.1$, and dropout rate ranging from $0.0$ to $0.75$. These experiments did not yield any notable trend and in turn have not included these results in our manuscript. However, any model weights and log files are available via request for reproduciblility. For the following Sections \ref{sec:results} and \ref{sec:experiments}, we utilize the hyperparameters which result in the top performing model for each feature type.\\

Our dataset and code are undergoing an IP review and are expected to be released to the public.

\section{Results} \label{sec:results}
We evaluate classifier performance along three dimensions: 
\begin{enumerate}
 \item[4.1] Model Comparison
 \item[4.2] Feature Comparison
 \item[4.3] Task Comparison
\end{enumerate}

For model comparison, we first compare our Parallel and Cascade architectures as seen in  Figure~\ref{fig:comparison}. Our goal is to provide unbiased data and to learn the scenarios under which each model performs better or worse than the other. We hypothesize that the Cascade model will demonstrate better performance on the misfire fault detection task over the Parallel approach. Next, with the feature comparison subsection, we illustrate the impact the choice of audio feature has on model accuracy. Finally, we delve into task comparison to better understand how our top performing models are classifying each of the four main attributes: fuel type, engine configuration, cylinder count, and aspiration type, as well these labels' contribution to misfire fault detection performance. We include evaluation metrics (precision/recall) and confusion matrices to provide additional insight into the true value of our approaches.

\subsection{Model Comparison} \label{sec:model-comparison}
Table \ref{tab:model-val} compares the validation set accuracy for each feature type between the Na{\"i}ve, Parallel and Cascade models. We first note that both the Parallel and Cascade models significantly outperform the Na{\"i}ve model in both attribute and misfire prediction. Next we find that the Cascade model demonstrate minor improvements or degradation over the Cascade model for attributes across our feature types. Finally, the Cascade model outperforms the Parallel model across every feature type for misfire prediction. Specifically, we observe 7.4\%, 4.2\%, 0.7\%, 12.3\%, 2.7\% validation set accuracy improvements for FFT, MFCC, spectrogram, waveform, and wavelet model performance respectively.\\

In exploring the test set results in Table~\ref{tab:model-test}, we observe similar patterns. With respect to the Na{\"i}ve baseline, it appears that the Parallel and Cascade models plateau for attributes. However, for misfire prediction the Cascade model outperforms both the Na{\"i}ve and Parallel baselines across every feature type. Specifically, when comparing Parallel and Cascade model test set accuracy on the misfire task, we observe 4.0\%, 1.5\%, 1.0\%, 8.8\%, 1.7\% improvements for FFT, MFCC, spectrogram, waveform, and wavelet model performance respectively.\\

\textbf{The major takeaway: the Cascade model outperforms all baselines on misfire prediction. These findings show we have achieved our goal of improving fault prediction through cascading of vehicle attributes.}.

\begin{table}[H]
\caption{Model comparison for validation set accuracy. We observe for the four attribute prediction tasks, Parallel and Cascade achieve similar performance whereas for the misfire task Cascade outperforms Parallel across all feature types.}
\label{tab:model-val}
\begin{center}
\begin{tabular}{|c|c|c|c|c|c|c|} \hline
\textbf{Model} & \textbf{Feature} & \textbf{Fuel} & \textbf{Config} & \textbf{Cyl} & \textbf{Turbo} & \textbf{Misfire} \\ \hline
Na{\"i}ve & -- & 86.7\% & 86.6\% & 74.9\% & 63.5\% & 77.2\% \\
\midrule
Parallel & FFT & 90.1\% & \textbf{89.9\%} & 78.2\% & \textbf{80.0\%} & 84.4\% \\ 
Cascade & FFT & \textbf{90.4\%} & \textbf{89.9\%} & \textbf{81.4\%} & 77.5\% & \textbf{91.8\%} \\ 
\midrule
Parallel & MFCC & \textbf{96.0\%} & \textbf{94.0\%} & 93.6\% & \textbf{92.9\%} & 89.4\% \\ 
Cascade & MFCC & 95.6\% & 93.7\% & \textbf{94.0\%} & \textbf{92.8\%} & \textbf{93.6\%} \\
\midrule
Parallel & Spectrogram & \textbf{87.7\%} & \textbf{86.7\%} & 75.9\% & \textbf{80.0\%} & 85.8\% \\ 
Cascade & Spectrogram & 87.4\% & \textbf{86.6\%} & \textbf{76.2\%} & 78.4\% & \textbf{86.5\%} \\
\midrule
Parallel & Waveform & \textbf{95.2\%} & \textbf{93.5\%} & 90.1\% & 90.5\% & 81.1\% \\ 
Cascade & Waveform & 94.9\% & 92.0\% & \textbf{90.8\%} & \textbf{92.0\%} & \textbf{93.4\%} \\
\midrule
Parallel & Wavelets & 85.9\% & 69.2\% & 59.5\% & 80.5\% & 86.7\% \\
Cascade & Wavelets & \textbf{86.6\%} & \textbf{76.4\%} & \textbf{67.9\%} & \textbf{81.4\%} & \textbf{89.4\%} \\
\hline
\end{tabular}
\end{center}
\end{table}

\begin{table}[H]
\caption{Model comparison for test set accuracy. We again observe relatively similar performance across attributes between Parallel and Cascade, while the Cascade model consistently improves upon the Parallel model for misfire prediction.}
\label{tab:model-test}
\begin{center}
\begin{tabular}{|c|c|c|c|c|c|c|} \hline
\textbf{Model} & \textbf{Feature} & \textbf{Fuel} & \textbf{Config} & \textbf{Cyl} & \textbf{Turbo} & \textbf{Misfire} \\ \hline
Na{\"i}ve & -- & 81.7\% & 78.3\% & 68.9\% & 66.8\% & 79.0\% \\
\midrule
Parallel & FFT & 78.8\% & \textbf{78.3\%} & \textbf{67.7\%} & \textbf{70.6\%} & 79.5\% \\ 
Cascade & FFT & \textbf{81.7\%} & 77.8\% & 68.1\% & 66.5\% & \textbf{83.5\%} \\
\midrule
Parallel & MFCC & \textbf{84.7\%} & \textbf{78.3\%} & \textbf{72.0\%} & \textbf{73.0\%} & 85.3\% \\ 
Cascade & MFCC & 81.6\% & \textbf{78.2\%} & 67.3\% & 70.1\% & \textbf{86.8\%} \\ 
\midrule
Parallel & Spectrogram & 77.0\% & \textbf{78.3\%} & \textbf{68.9\%} & 65.9\% & 84.9\% \\ 
Cascade & Spectrogram & \textbf{81.7\%} & \textbf{78.3\%} & \textbf{68.9\%} & \textbf{67.2\%} & \textbf{85.9\%} \\
\midrule
Parallel & Waveform & 73.3\% & 75.2\% & \textbf{64.7\%} & \textbf{67.7\%} & 70.8\% \\ 
Cascade & Waveform & \textbf{82.0\%} & \textbf{78.3\%} & 52.5\% & 67.3\% & \textbf{79.6\%} \\
\midrule
Parallel & Wavelets & \textbf{79.9\%} & \textbf{63.4\%} & 52.7\% & \textbf{69.0\%} & 81.4\% \\ 
Cascade & Wavelets & 78.4\% & 62.1\% & \textbf{53.1\%} & 66.8\% & \textbf{83.1\%} \\
\hline
\end{tabular}
\end{center}
\end{table}

\subsection{Feature Comparison}
We explore model performance for five feature types (FFT, MFCC, spectrogram, wavelets, waveform). FFT, wavelets, and waveform share the exact same model architecture, whereas MFCC and spectrogram each use a unique model architecture with differing amounts of layers and kernel sizes to reflect the feature dimensionality.\\ 

In investigating validation set performance from Table~\ref{tab:feature-val}, we see that MFCC features outperform all other feature types in both the Parallel and Cascade models. When exploring the test set performance from Table \ref{tab:feature-test}, we again find MFCC is again the top performing feature for the Parallel and Cascade models. We also note that the models with spectrogram features achieve second place for both Parallel and Cascade on the test set. This shows that the models which utilize 2D convolution are able to find more generalizable patterns from the train set to the test set. The final pattern we can observe is the poor performance of the raw waveform relative to the other feature types. This demonstrates the value of our experiments which analyze many different types of audio features. 

\begin{table}[H]
\caption{Feature comparison for validation set accuracy. MFCC features perform the strongest across all attributes and misfire for both the Parallel and Cascade models.}
\label{tab:feature-val}
\begin{center}
\begin{tabular}{|c|c|c|c|c|c|c|} \hline
\textbf{Model} & \textbf{Feature} & \textbf{Fuel} & \textbf{Config} & \textbf{Cyl} & \textbf{Turbo} & \textbf{Misfire} \\ \hline
Parallel & MFCC & \textbf{96.0\%} & \textbf{94.0\%} & \textbf{93.6\%} & \textbf{92.9\%} & \textbf{89.4\%} \\
Parallel & Wavelets & 85.9\% & 69.2\% & 59.5\% & 80.5\% & 86.7\% \\ 
Parallel & Spectrogram & 87.7\% & 86.7\% & 75.9\% & 80.0\% & 85.8\% \\ 
Parallel & FFT & 90.1\% & 89.9\% & 78.2\% & 80.0\% & 84.4\% \\
Parallel & Waveform & 95.2\% & 93.5\% & 90.1\% & 90.5\% & 81.1\% \\
\midrule
Cascade & MFCC & \textbf{95.6\%} & \textbf{93.7\%} & \textbf{94.0\%} & \textbf{92.8\%} & \textbf{93.6\%} \\ 
Cascade & Waveform & 94.9\% & 92.0\% & 90.8\% & 92.0\% & 93.4\% \\ 
Cascade & FFT & 90.4\% & 89.9\% & 81.4\% & 77.5\% & 91.8\% \\ 
Cascade & Wavelets & 86.6\% & 76.4\% & 67.9\% & 81.4\% & 89.4\% \\
Cascade & Spectrogram & 87.4\% & 86.6\% & 76.2\% & 78.4\% & 86.5\% \\ 
\hline
\end{tabular}
\end{center}
\end{table}

\begin{table}[H]
\caption{Feature comparison for test set accuracy. For the Parallel model MFCC features translate their strong validation performance to test set generalizability. However, the Cascade model better utilizes FFT and spectrogram features for attributes and spectrogram and wavelet features for misfire.}
\label{tab:feature-test}
\begin{center}
\begin{tabular}{|c|c|c|c|c|c|c|} \hline
\textbf{Model} & \textbf{Feature} & \textbf{Fuel} & \textbf{Config} & \textbf{Cyl} & \textbf{Turbo} & \textbf{Misfire} \\ \hline
Parallel & MFCC & \textbf{84.7\%} & \textbf{78.3\%} & \textbf{72.0\%} & \textbf{73.0\%} & \textbf{85.3\%} \\ 
Parallel & Spectrogram & 77.0\% & \textbf{78.3\%} & 68.9\% & 65.9\% & 84.9\% \\
Parallel & Wavelets & 79.9\% & 63.4\% & 52.7\% & 69.0\% & 81.4\% \\
Parallel & FFT & 78.8\% & 78.3\% & 67.7\% & 70.6\% & 79.5\% \\ 
Parallel & Waveform & 73.3\% & 75.2\% & 64.7\% & 67.7\% & 70.8\% \\
\midrule
Cascade & MFCC & 81.6\% & \textbf{78.2\%} & 67.3\% & \textbf{70.1\%} & \textbf{86.8\%} \\ 
Cascade & Spectrogram & 81.7\% & \textbf{78.3\%} & \textbf{68.9\%} & 67.2\% & 85.9\% \\
Cascade & FFT & 81.7\% & 77.8\% & 68.1\% & 66.5\% & 83.5\% \\
Cascade & Wavelets & 78.4\% & 62.1\% & 53.1\% & 66.8\% & 83.1\% \\
Cascade & Waveform & \textbf{82.0\%} & \textbf{78.3\%} & 52.5\% & 67.3\% & 79.6\% \\
\hline
\end{tabular}
\end{center}
\end{table}

\subsection{Task Comparison}
We explore the confusion matrices for each of the four attribute and the misfire prediction tasks:
\begin{enumerate}
 \item[4.3.1] Fuel Type
 \item[4.3.2] Engine Configuration
 \item[4.3.3] Cylinder Count
 \item[4.3.4] Aspiration Type
 \item[4.3.5] Misfire Detection
\end{enumerate}
These results use both the Parallel and Cascade Models with the top performing feature type, MFCCs. 
\newpage
\subsubsection{Fuel Type}
With our first prediction task, fuel type, in Table~\ref{tab:fuel-conf-val} we observe very high validation set accuracy ($98\%$) on the gasoline class. However, due to class imbalance, the Diesel class validation accuracy was lower ($84\%$). This pattern follows in Table~\ref{tab:fuel-conf-test} with strong test set accuracy on gasoline ($97\%$). Unfortunately, there was a significant degradation from validation to test set accuracy on the underrepresented Diesel class ($32\%$). This is further represented in the precision and recall metrics in Table \ref{tab:fuel-metrics} with Cascade achieving 90.4\% precision and 89.1\% recall on the validation set while degrading to 63.5\% precision and 67.6\% recall on the test set.\\ 

Of note, the Cascade model outperformed the Parallel model on the underrepresented Diesel class in both validation and test sets. We hypothesize the Cascade model is able to perform better on the underrepresented class due to its separate stage CNNs, allowing the network more distinctly differentiate samples, particularly those that are abnormal and may be more difficult to classify.

\begin{table}[H]
\caption{Fuel Type validation and test set metrics using MFCC features.}
\label{tab:fuel-metrics}
\begin{center}
\begin{tabular}{|c|c|c|c|c|} \hline
\textbf{Model} & \textbf{Set} &\textbf{Accuracy} & \textbf{Precision} & \textbf{Recall}  \\ \hline
Parallel & Validation & \textbf{96.0\%} & 88.1\% & \textbf{90.4\%}  \\ 
Cascade & Validation & 95.6\% & \textbf{90.4\%} & 89.1\%  \\
\midrule
Parallel & Test & \textbf{84.7\%} & 60.5\% & 66.5\%  \\ 
Cascade & Test & 81.6\% & \textbf{63.5\%} & \textbf{67.6\%}  \\
\hline
\end{tabular}
\end{center}
\end{table}

{
\makegapedcells
\begin{table}[H]
\centering
\caption{Fuel Type validation confusion matrix for Parallel (left) and Cascade (right) using MFCC features.}
\label{tab:fuel-conf-val}
\begin{tabular}{cc|cc}
    \multicolumn{2}{c}{}
    &   \multicolumn{2}{c}{Predicted} \\
    &       &   Diesel &   Gasoline              \\ 
    \cline{2-4}
    \multirow{2}{*}{\rotatebox[origin=c]{90}{Actual}}
    & Diesel & \textbf{580 (81.8\%)}& \textcolor{black}{129 (18.2\%)}\\
    & Gasoline & \textcolor{black}{83 (1.8\%)}& \textbf{4525 (98.2\%)}\\
    \cline{2-4}
\end{tabular}
\begin{tabular}{cc|cc}
    \multicolumn{2}{c}{}
    &   \multicolumn{2}{c}{Predicted} \\
    &       &   Diesel &   Gasoline              \\ 
    \cline{2-4}
    \multirow{2}{*}{\rotatebox[origin=c]{90}{Actual}}
    & Diesel & \textbf{594 (83.8\%)}& \textcolor{black}{115 (16.2\%)}\\
    & Gasoline & \textcolor{black}{142 (3.1\%)}& \textbf{4466 (96.9\%)}\\
    \cline{2-4}
\end{tabular}
\end{table}
}

{
\makegapedcells
\begin{table}[H]
\centering
\caption{Fuel Type test confusion matrix for Parallel (left) and Cascade (right) using MFCC features.}
\label{tab:fuel-conf-test}
\begin{tabular}{cc|cc}
    \multicolumn{2}{c}{}
    &   \multicolumn{2}{c}{Predicted} \\
    &       &   Diesel &   Gasoline              \\ 
    \cline{2-4}
    \multirow{2}{*}{\rotatebox[origin=c]{90}{Actual}}
    & Diesel & \textbf{43 (24.7\%)}& \textcolor{black}{131 (75.3\%)}\\
    & Gasoline & \textcolor{black}{22 (2.8\%)}& \textbf{755 (97.1\%)}\\
    \cline{2-4}
\end{tabular}
\begin{tabular}{cc|cc}
    \multicolumn{2}{c}{}
    &   \multicolumn{2}{c}{Predicted} \\
    &       &   Diesel &   Gasoline              \\ 
    \cline{2-4}
    \multirow{2}{*}{\rotatebox[origin=c]{90}{Actual}}
    & Diesel & \textbf{56 (32.2\%)}& \textcolor{black}{118 (67.8\%)}\\
    & Gasoline & \textcolor{black}{58 (7.5\%)}& \textbf{719 (92.5\%)}\\
    \cline{2-4}
\end{tabular}
\end{table}
}
\subsubsection{Engine Configuration}
Our next prediction task focused on engine configuration: Flat (Boxer), Inline, or Vee. Validation and test set results are seen in Table \ref{tab:config-conf-val} and Table \ref{tab:config-conf-test}, respectively. The first observation is the lack of representation with the rare engine configuration class of Flat, which therefore did not lead to the identification of a meaningful informative feature pattern. \\

Similarly to Fuel Type, our models show strong performance on the most well-represented class, in this case, Inline with $98\%$ and $93\%$ validation and test set accuracy respectively. The secondary class V configuration achieves underwhelming validation accuracy of $67\%$ with again degradation down to $26\%$ on the test set. Further elucidated in Table \ref{tab:config-metrics} where the Cascade model precision and recall degrades from 73.2\% and 79.9\% to 49.8\% and 53.7\% from validation to test set. 

\begin{table}[H]
\caption{Engine Configuration validation and test set metrics using MFCC features.}
\label{tab:config-metrics}
\begin{center}
\begin{tabular}{|c|c|c|c|c|} \hline
\textbf{Model} & \textbf{Set} &\textbf{Accuracy} & \textbf{Precision} & \textbf{Recall}  \\ \hline
Parallel & Validation & \textbf{94.0\%} & \textbf{73.5\%} & 79.6\%  \\ 
Cascade & Validation & 93.7\% & 73.2\% & \textbf{79.9\%}  \\
\midrule
Parallel & Test & \textbf{78.3\%} & 47.2\% & 49.6\%  \\ 
Cascade & Test & \textbf{78.2\%} & \textbf{49.8\%} & \textbf{53.7\%}  \\
\hline
\end{tabular}
\end{center}
\end{table}

{
\makegapedcells
\begin{table}[H]
\centering
\caption{Engine configuration validation confusion matrix for Parallel (top) and Cascade (bottom) using MFCC features.}
\label{tab:config-conf-val}
\begin{tabular}{cc|ccc}
    \multicolumn{2}{c}{}
    &   \multicolumn{3}{c}{Predicted} \\
    &       &   Flat & Inline & V              \\ 
    \cline{2-5}
    \multirow{4}{*}{\rotatebox[origin=c]{90}{Actual}}
    & Flat & \textbf{0 (0.0\%)} & \textcolor{black}{7 (53.9\%)} & \textcolor{black}{6 (46.1\%)} \\
    & Inline & \textcolor{black}{0 (0.0\%)}& \textbf{4499 (98.4\%)} & \textcolor{black}{73 (1.6\%)}\\
    & V & \textcolor{black}{0 (0.0\%)} & \textcolor{black}{229 (33.0\%)}& \textbf{464 (67.0\%)}\\
    \cline{2-5}
\end{tabular}
\begin{tabular}{cc|ccc}
    \multicolumn{2}{c}{}
    &   \multicolumn{3}{c}{Predicted} \\
    &       &   Flat & Inline & V             \\ 
    \cline{2-5}
    \multirow{4}{*}{\rotatebox[origin=c]{90}{Actual}}
    & Flat & \textbf{0 (0.0\%)} & \textcolor{black}{7 (53.9\%)} & \textcolor{black}{6 (46.1\%)} \\
    & Inline & \textcolor{black}{0 (0.0\%)}& \textbf{4474(97.9\%)} & \textcolor{black}{98 (2.1\%)}\\
    & V & \textcolor{black}{0 (0.0\%)} & \textcolor{black}{250 (36.1\%)}& \textbf{443 (63.9\%)}\\
    \cline{2-5}
\end{tabular}
\end{table}
}

{
\makegapedcells
\begin{table}[H]
\centering
\caption{Engine Configuration test confusion matrix for Parallel (left) and Cascade (right) using MFCC features.}
\label{tab:config-conf-test}
\begin{tabular}{cc|ccc}
    \multicolumn{2}{c}{}
    &   \multicolumn{3}{c}{Predicted} \\
    &       &   Flat & Inline & V             \\ 
    \cline{2-5}
    \multirow{4}{*}{\rotatebox[origin=c]{90}{Actual}}
    & Flat & \textbf{0 (0.0\%)} & \textcolor{black}{0 (0.0\%)} & \textcolor{black}{2 (100.0\%)} \\
    & Inline & \textcolor{black}{0 (0.0\%)} & \textbf{696 (93.4\%)} & \textcolor{black}{49 (6.6\%)}\\
    & V & \textcolor{black}{0 (0.0\%)} & \textcolor{black}{154 (75.5\%)}& \textbf{50 (25.5\%)}\\
    \cline{2-5}
\end{tabular}
\begin{tabular}{cc|ccc}
    \multicolumn{2}{c}{}
    &  \multicolumn{3}{c}{Predicted}\\
    &       &   Flat & Inline & V              \\ 
    \cline{2-5}
    \multirow{4}{*}{\rotatebox[origin=c]{90}{Actual}}
    & Flat & \textbf{0 (0.0\%)} & \textcolor{black}{0 (0.0\%)} & \textcolor{black}{2 (100.0\%)} \\
    & Inline & \textcolor{black}{0 (0.0\%)} & \textbf{693 (93.0\%)} & \textcolor{black}{52 (7.0\%)}\\
    & V & \textcolor{black}{0 (0.0\%)} & \textcolor{black}{173 (84.8\%)}& \textbf{31 (15.2\%)}\\
    \cline{2-5}
\end{tabular}
\end{table}
}

\newpage
\subsubsection{Cylinder Count}
The next prediction task we explore confusion matrices for is cylinder counts in the range from two to 8, seen in Tables \ref{tab:cyl-conf-val} and \ref{tab:cyl-conf-test}. Although we know there is an under-representation of classes 2, 3, and 5 in the validation and test set from Figures \ref{fig:label-val} and \ref{fig:label-test}, we observe strong validation performance on classes 2 ($68\%$), 3 ($68\%$) and five ($96\%$) (perhaps because of potential for imbalance more likely to occur in odd-numbered cylinder count engines). For the better-represented classes of six and 8 cylinders, we find $78\%$ and $89\%$ accuracy, with the most well-represented class of four cylinders achieving $99\%$ validation accuracy.\\

Again, following the trend of test set degradation from the fuel and config tasks, we observe only good performance ($93\%$) on the four-cylinder class. Given the more challenging prediction task across six labels, we observe our largest degradation in terms of precision and recall as seen in Table \ref{tab:cyl-metrics}. Specifically, the Cascade model's strong validation precision of 77.9\% and recall of 87.0\% significantly decreases to 21.9\% and 26.4\%. With the weak performance on both validation and test for multiple prediction tasks, this motivates two directions of future work. The first would be building a larger and more balanced training set. The second direction would be modifying the training process with hard negative mining or label weighting.

\begin{table}[H]
\caption{Cylinder Count validation and test set metrics using MFCC features.}
\label{tab:cyl-metrics}
\begin{center}
\begin{tabular}{|c|c|c|c|c|} \hline
\textbf{Model} & \textbf{Set} &\textbf{Accuracy} & \textbf{Precision} & \textbf{Recall}  \\ \hline
Parallel & Validation & 93.6\% & 77.0\% & 85.0\%  \\ 
Cascade & Validation & \textbf{94.0\%} & \textbf{77.9\%} & \textbf{87.0\%}  \\
\midrule
Parallel & Test & \textbf{72.0\%} & \textbf{26.4\%} & \textbf{31.9\%}  \\ 
Cascade & Test & 67.3\% & 21.9\% & 26.4\%  \\
\hline
\end{tabular}
\end{center}
\end{table}

{
\makegapedcells
\begin{table}[H]
\centering
\caption{Cylinder Count validation confusion matrix for Parallel (top) and Cascade (bottom) using MFCC features.}
\label{tab:cyl-conf-val}
\begin{tabular}{cc|cccccc}
    \multicolumn{2}{c}{}
    &   \multicolumn{6}{c}{Predicted} \\
    &       &   2 & 3 & 4 & 5 & 6 & 8             \\ 
    \cline{2-8}
    \multirow{9}{*}{\rotatebox[origin=c]{90}{Actual}}
& 2 & \textbf{25 (67.6\%)} & \textcolor{black}{0 (0.0\%)} & \textcolor{black}{12 (32.4\%)} & \textcolor{black}{0 (0.0\%)} & \textcolor{black}{0 (0.0\%)} & \textcolor{black}{0 (0.0\%)} \\
& 3 & \textcolor{black}{0 (0.0\%)} & \textbf{75 (67.6\%)} & \textcolor{black}{30 (27.0\%)} & \textcolor{black}{0 (0.0\%)} & \textcolor{black}{6 (5.4\%)} & \textcolor{black}{0 (0.0\%)} \\
& 4 & \textcolor{black}{0 (0.0\%)} & \textcolor{black}{0 (0.0\%)} & \textbf{3833 (98.9\%)} & \textcolor{black}{2 (0.1\%)} & \textcolor{black}{39 (1.0\%)} & \textcolor{black}{0 (0.0\%)} \\
& 5 & \textcolor{black}{0 (0.0\%)} & \textcolor{black}{0 (0.0\%)} & \textcolor{black}{3 (6.2\%)} & \textbf{45 (93.8\%)} & \textcolor{black}{0 (0.0\%)} & \textcolor{black}{0 (0.0\%)} \\
& 6 & \textcolor{black}{1 (0.1\%)} & \textcolor{black}{0 (0.0\%)} & \textcolor{black}{213 (22.3\%)} & \textcolor{black}{0 (0.0\%)} & \textbf{740 (77.5\%)} & \textcolor{black}{1 (0.1\%)} \\
& 8 & \textcolor{black}{0 (0.0\%)} & \textcolor{black}{0 (0.0\%)} & \textcolor{black}{27 (10.9\%)} & \textcolor{black}{0 (0.0\%)} & \textcolor{black}{0 (0.0\%)} & \textbf{221 (89.1\%)}\\
    \cline{2-8}
\end{tabular}
\begin{tabular}{cc|cccccc}
    \multicolumn{2}{c}{}
    &   \multicolumn{6}{c}{Predicted} \\
    &       &   2 & 3 & 4 & 5 & 6 & 8             \\ 
    \cline{2-8}
    \multirow{9}{*}{\rotatebox[origin=c]{90}{Actual}}
& 2 & \textbf{24 (64.9\%)} & \textcolor{black}{0 (0.0\%)} & \textcolor{black}{4 (10.8\%)} & \textcolor{black}{0 (0.0\%)} & \textcolor{black}{9 (24.3\%)} & \textcolor{black}{0 (0.0\%)} \\
& 3 & \textcolor{black}{0 (0.0\%)} & \textbf{69 (62.2\%)} & \textcolor{black}{38 (34.2\%)} & \textcolor{black}{0 (0.0\%)} & \textcolor{black}{4 (3.6\%)} & \textcolor{black}{0 (0.0\%)} \\
& 4 & \textcolor{black}{0 (0.0\%)} & \textcolor{black}{0 (0.0\%)} & \textbf{3812 (98.4\%)} & \textcolor{black}{0 (0.0\%)} & \textcolor{black}{59 (1.5\%)} & \textcolor{black}{3 (0.1\%)} \\
& 5 & \textcolor{black}{0 (0.0\%)} & \textcolor{black}{0 (0.0\%)} & \textcolor{black}{2 (4.2\%)} & \textbf{46 (95.8\%)} & \textcolor{black}{0 (0.0\%)} & \textcolor{black}{0 (0.0\%)} \\
& 6 & \textcolor{black}{4 (0.4\%)} & \textcolor{black}{0 (0.0\%)} & \textcolor{black}{226 (23.7\%)} & \textcolor{black}{0 (0.0\%)} & \textbf{721 (75.5\%)} & \textcolor{black}{4 (0.4\%)} \\
& 8 & \textcolor{black}{0 (0.0\%)} & \textcolor{black}{0 (0.0\%)} & \textcolor{black}{38 (15.3\%)} & \textcolor{black}{0 (0.0\%)} & \textcolor{black}{1 (0.4\%)} & \textbf{209 (84.3\%)}\\
    \cline{2-8}
\end{tabular}
\end{table}
}

{
\makegapedcells
\begin{table}[H]
\centering
\caption{Cylinder Count test confusion matrix for Parallel (top) and Cascade (bottom) using MFCC features.}
\label{tab:cyl-conf-test}
\begin{tabular}{cc|cccccc}
    \multicolumn{2}{c}{}
    &   \multicolumn{6}{c}{Predicted} \\
    &       &  2 & 3 & 4 & 5 & 6 & 8             \\ 
    \cline{2-8}
    \multirow{9}{*}{\rotatebox[origin=c]{90}{Actual}}
& 2 & \textbf{0 (0.0\%)} & \textcolor{black}{0 (0.0\%)} & \textcolor{black}{19 (100.0\%)} & \textcolor{black}{0 (0.0\%)} & \textcolor{black}{0 (0.0\%)} & \textcolor{black}{0 (0.0\%)} \\
& 3 & \textcolor{black}{0 (0.0\%)} & \textbf{0 (0.0\%)} & \textcolor{black}{30 (100.0\%)} & \textcolor{black}{0 (0.0\%)} & \textcolor{black}{0 (0.0\%)} & \textcolor{black}{0 (0.0\%)} \\
& 4 & \textcolor{black}{0 (0.0\%)} & \textcolor{black}{0 (0.0\%)} & \textbf{611 (93.3\%)} & \textcolor{black}{3 (0.5\%)} & \textcolor{black}{32 (4.9\%)} & \textcolor{black}{9 (1.4\%)} \\
& 5 & \textcolor{black}{0 (0.0\%)} & \textcolor{black}{0 (0.0\%)} & \textcolor{black}{9 (100.0\%)} & \textbf{0 (0.0\%)} & \textcolor{black}{0 (0.0\%)} & \textcolor{black}{0 (0.0\%)} \\
& 6 & \textcolor{black}{1 (0.1\%)} & \textcolor{black}{0 (0.0\%)} & \textcolor{black}{132 (75.4\%)} & \textcolor{black}{0 (0.0\%)} & \textbf{43 (24.6\%)} & \textcolor{black}{0 (0.0\%)} \\
& 8 & \textcolor{black}{0 (0.0\%)} & \textcolor{black}{0 (0.0\%)} & \textcolor{black}{30 (47.6\%)} & \textcolor{black}{0 (0.0\%)} & \textcolor{black}{0 (0.0\%)} & \textbf{33 (52.4\%)}\\
    \cline{2-8}
\end{tabular}
\begin{tabular}{cc|cccccc}
    \multicolumn{2}{c}{}
    &  \multicolumn{6}{c}{Predicted}\\
    &       &   2 & 3 & 4 & 5 & 6 & 8              \\ 
    \cline{2-8}
    \multirow{9}{*}{\rotatebox[origin=c]{90}{Actual}}
& 2 & \textbf{2 (10.5\%)} & \textcolor{black}{0 (0.0\%)} & \textcolor{black}{17 (89.5\%)} & \textcolor{black}{0 (0.0\%)} & \textcolor{black}{0 (0.0\%)} & \textcolor{black}{0 (0.0\%)} \\
& 3 & \textcolor{black}{0 (0.0\%)} & \textbf{0 (0.0\%)} & \textcolor{black}{30 (100.0\%)} & \textcolor{black}{0 (0.0\%)} & \textcolor{black}{0 (0.0\%)} & \textcolor{black}{0 (0.0\%)} \\
& 4 & \textcolor{black}{17 (2.6\%)} & \textcolor{black}{3 (0.5\%)} & \textbf{569 (86.9\%)} & \textcolor{black}{2 (0.3\%)} & \textcolor{black}{54 (8.2\%)} & \textcolor{black}{10 (1.5\%)} \\
& 5 & \textcolor{black}{0 (0.0\%)} & \textcolor{black}{0 (0.0\%)} & \textcolor{black}{9 (100.0\%)} & \textbf{0 (0.0\%)} & \textcolor{black}{0 (0.0\%)} & \textcolor{black}{0 (0.0\%)} \\
& 6 & \textcolor{black}{1 (0.1\%)} & \textcolor{black}{0 (0.0\%)} & \textcolor{black}{153 (87.4\%)} & \textcolor{black}{0 (0.0\%)} & \textbf{22 (12.6\%)} & \textcolor{black}{0 (0.0\%)} \\
& 8 & \textcolor{black}{0 (0.0\%)} & \textcolor{black}{0 (0.0\%)} & \textcolor{black}{53 (84.1\%)} & \textcolor{black}{0 (0.0\%)} & \textcolor{black}{0 (0.0\%)} & \textbf{10 (15.9\%)}\\
    \cline{2-8}
\end{tabular}
\end{table}
}

\subsubsection{Aspiration Type}
We now explore the aspiration type task which looks to predict normally aspirated or turbocharged engine variants as seen in Tables \ref{tab:turbo-conf-val} and \ref{tab:turbo-conf-test}. Both the Parallel and Cascade models demonstrate strong performance on both classes with the validation set $94\%$ for normally-aspirated and $91\%$ for turbocharged. The strong performance on the normal label translates to the test set with $91\%$ but degrades to $41\%$ for turbocharged. This poor performance on the underrepresented label affects the low precision and recall (63.4\% / 66.9\%) on the test set for the Cascade model as seen in Table \ref{tab:turbo-metrics}.\\

It is also worth noting that the Cascade model performs better on the underrepresented Turbocharged class similarly to its performance on the underrepresented Diesel class from the fuel task. As we mentioned earlier, the Cascade model having two distinct stages may allow it to specialize on underrepresented classes.

\begin{table}[H]
\caption{Aspiration Type validation and test set metrics using MFCC features.}
\label{tab:turbo-metrics}
\begin{center}
\begin{tabular}{|c|c|c|c|c|} \hline
\textbf{Model} & \textbf{Set} &\textbf{Accuracy} & \textbf{Precision} & \textbf{Recall}  \\ \hline
Parallel & Validation & \textbf{92.9\%} & 91.2\% & \textbf{92.0\%}  \\ 
Cascade & Validation & \textbf{92.8\%} & \textbf{92.1\%} & 91.5\%  \\
\midrule
Parallel & Test & \textbf{73.0\%} & \textbf{65.5\%} & \textbf{74.7\%}  \\ 
Cascade & Test & 70.1\% & 63.4\% & 66.9\%  \\
\hline
\end{tabular}
\end{center}
\end{table}

{
\makegapedcells
\begin{table}[H]
\centering
\caption{Aspiration Type validation confusion matrix for Parallel (left) and Cascade (right) using MFCC features.}
\label{tab:turbo-conf-val}
\begin{tabular}{cc|cc}
    \multicolumn{2}{c}{}
    &   \multicolumn{2}{c}{Predicted} \\
    &       &   Normal &   Turbocharged              \\ 
    \cline{2-4}
    \multirow{2}{*}{\rotatebox[origin=c]{90}{Actual}}
    & Normal & \textbf{3126 (94.5\%)}& \textcolor{black}{183 (5.5\%)}\\
    & Turbocharged & \textcolor{black}{188 (9.9\%)}& \textbf{1712 (90.1\%)}\\
    \cline{2-4}
\end{tabular}
\begin{tabular}{cc|cc}
    \multicolumn{2}{c}{}
    &   \multicolumn{2}{c}{Predicted} \\
    &       &   Normal &   Turbocharged              \\ 
    \cline{2-4}
    \multirow{2}{*}{\rotatebox[origin=c]{90}{Actual}}
    & Normal & \textbf{3082 (93.1\%)}& \textcolor{black}{227 (6.9\%)}\\
    & Turbocharged & \textcolor{black}{171 (9.0\%)}& \textbf{1729 (91.0\%)}\\
    \cline{2-4}
\end{tabular}
\end{table}
}

{
\makegapedcells
\begin{table}[H]
\centering
\caption{Aspiration Type test confusion matrix for Parallel (left) and Cascade (right) using MFCC features.}
\label{tab:turbo-conf-test}
\begin{tabular}{cc|cc}
    \multicolumn{2}{c}{}
    &   \multicolumn{2}{c}{Predicted} \\
    &       &   Normal &   Turbocharged              \\ 
    \cline{2-4}
    \multirow{2}{*}{\rotatebox[origin=c]{90}{Actual}}
    & Normal & \textbf{579 (91.2\%)}& \textcolor{black}{56 (8.8\%)}\\
    & Turbocharged & \textcolor{black}{203 (64.2\%)}& \textbf{113 (35.8\%)}\\
    \cline{2-4}
\end{tabular}
\begin{tabular}{cc|cc}
    \multicolumn{2}{c}{}
    &   \multicolumn{2}{c}{Predicted} \\
    &       &   Normal &   Turbocharged              \\ 
    \cline{2-4}
    \multirow{2}{*}{\rotatebox[origin=c]{90}{Actual}}
    & Normal & \textbf{548 (86.3\%)}& \textcolor{black}{87 (13.7\%)}\\
    & Turbocharged & \textcolor{black}{188 (59.5\%)}& \textbf{128 (40.5\%)}\\
    \cline{2-4}
\end{tabular}
\end{table}
}

\subsubsection{Misfire Detection}
Our final prediction task is misfire detection, which looks to predict engine status as normal or abnormal due to an engine misfire (incomplete combustion) which in our set are found only in gasoline engine samples. Results appear in Tables \ref{tab:misfire-conf-val} and \ref{tab:misfire-conf-test}. As discussed in Section~\ref{sec:model-comparison}, the Cascade model fulfills its purpose in our proof-of-concept by improving the misfire detection performance on both the validation and test set relative to the Parallel model. This is evidenced in Table \ref{tab:misfire-metrics} as the Cascade model achieves a margin over the Parallel model of 4.2\%, 5.4\% and 3.2\% for validation accuracy, precision, and recall, respectively. Additionally, in Table \ref{tab:misfire-metrics} we find a test set margin for the Cascade model over the Parallel model of 1.5\%, 1.3\% and 5.3\% for accuracy, precision, and recall, respectively.\\ 

We can further see the strength of the Cascade model with $96\%$ and $90\%$ accuracy on abnormal and normal classes, respectively. It is particularly interesting that the misfire task is the only task in our exploration where performance on the less represented class is better than the more represented class. However, as with all of our experiments, the models experience degradation on the test set, specifically decreasing to $38\%$ Abnormal accuracy, while increasing to $98\%$ Normal accuracy. In practice, this low false negative rate may be valuable for manufacturers as it makes vehicles seem more reliable than they are. However, there is still work to be done to improve the false positive rate such that this becomes a useful tool for mechanics aiming to maintain peak vehicle efficiency. 

\begin{table}[H]
\caption{Misfire Detection validation and test set metrics using MFCC features.}
\label{tab:misfire-metrics}
\begin{center}
\begin{tabular}{|c|c|c|c|c|} \hline
\textbf{Model} & \textbf{Set} &\textbf{Accuracy} & \textbf{Precision} & \textbf{Recall}  \\ \hline
Parallel & Validation & 89.4\% & 87.7\% & 82.9\%  \\ 
Cascade & Validation & \textbf{93.6\%} & \textbf{93.1\%} & \textbf{86.1\%}  \\
\midrule
Parallel & Test & 85.3\% & 68.3\% & 85.5\%  \\ 
Cascade & Test & \textbf{86.8\%} & \textbf{69.6\%} & \textbf{90.8\%}  \\
\hline
\end{tabular}
\end{center}
\end{table}

{
\makegapedcells
\begin{table}[H]
\centering
\caption{Misfire Detection validation confusion matrix for Parallel (left) and Cascade (right) using MFCC features.}
\label{tab:misfire-conf-val}
\begin{tabular}{cc|cc}
    \multicolumn{2}{c}{}
    &   \multicolumn{2}{c}{Predicted} \\
    &       &   Abnormal &   Normal              \\ 
    \cline{2-4}
    \multirow{2}{*}{\rotatebox[origin=c]{90}{Actual}}
    & Abnormal & \textbf{1160 (93.2\%)}& \textcolor{black}{84 (6.8\%)}\\
    & Normal & \textcolor{black}{490 (11.6\%)}& \textbf{3721 (88.4\%)}\\
    \cline{2-4}
\end{tabular}
\begin{tabular}{cc|cc}
    \multicolumn{2}{c}{}
    &   \multicolumn{2}{c}{Predicted} \\
    &       &   Abnormal &   Normal              \\ 
    \cline{2-4}
    \multirow{2}{*}{\rotatebox[origin=c]{90}{Actual}}
    & Abnormal & \textbf{1199 (96.4\%)}& \textcolor{black}{45 (3.6\%)}\\
    & Normal & \textcolor{black}{432 (10.3\%)}& \textbf{3779 (89.7\%)}\\
    \cline{2-4}
\end{tabular}
\end{table}
}

{
\makegapedcells
\begin{table}[H]
\centering
\caption{Misfire Detection test confusion matrix for Parallel (left) and Cascade (right) using MFCC features.}
\label{tab:misfire-conf-test}
\begin{tabular}{cc|cc}
    \multicolumn{2}{c}{}
    &   \multicolumn{2}{c}{Predicted} \\
    &       &   Abnormal &   Normal              \\ 
    \cline{2-4}
    \multirow{2}{*}{\rotatebox[origin=c]{90}{Actual}}
    & Abnormal & \textbf{76 (38.0\%)}& \textcolor{black}{124 (62.0\%)}\\
    & Normal & \textcolor{black}{15 (2.0\%)}& \textbf{736 (98.0\%)}\\
    \cline{2-4}
\end{tabular}
\begin{tabular}{cc|cc}
    \multicolumn{2}{c}{}
    &   \multicolumn{2}{c}{Predicted} \\
    &       &   Abnormal &   Normal              \\ 
    \cline{2-4}
    \multirow{2}{*}{\rotatebox[origin=c]{90}{Actual}}
    & Abnormal & \textbf{81 (40.5\%)}& \textcolor{black}{119 (59.5\%)}\\
    & Normal & \textcolor{black}{6 (0.8\%)}& \textbf{745 (99.2\%)}\\
    \cline{2-4}
\end{tabular}
\end{table}
}

\subsubsection*{Task Comparison Summary}
We examined the evaluation metrics and confusion matrices across four vehicle attribute tasks and one misfire detection task for the Parallel baseline and proposed Cascade model. Our goal with the cascading architecture is to better inform status prediction through cascading of vehicle characterization features. We achieve this goal through the Cascade model outperforming the Parallel baseline for validation and test set evaluation metrics and confusion matrices in Tables \ref{tab:misfire-metrics}-\ref{tab:misfire-conf-test}. The Cascade model achieves this margin while only observing minimal degradation or slight improvement on the attribute tasks.

\section{Experimental Studies} \label{sec:experiments}
To prove model robustness for previously unseen vehicles and vehicle variants, we conducted three ablation studies: 
\begin{enumerate}
 \item[5.1] Data Augmentation
 \item[5.2] Feature Fusion
 \item[5.3] YouTube Outsample
\end{enumerate}

As discussed in Section~\ref{sec:related-data-augmentation}, data augmentation is an essential component of modern deep learning systems. We first explore how our models perform on the test set when trained with and without data augmentation.\\ 

Another component of successful neural networks is well-tuned hyperparameters. In particular, learning rate is difficult to optimize. We therefore consider seven rates in both model the Parallel and Cascade models.\\ 

Finally, we built an additional test set to further consider our models outsample generalizability by using input data crowd-sourced from YouTube. Through these test set experiments, we demonstrate why this is not an effective approach to evaluating real-world model performance. In particular, training on crowdsourced data from YouTube is not well-suited to model creation that is more broadly generalizable.

\subsection{Data Augmentation} \label{sec:ablation-aug}
Our first ablation explores the difference of training on augmented samples vs. real samples on model performance for our Parallel and Cascade models as seen in Tables \ref{tab:aug-parallel-val}-\ref{tab:aug-cascade-test}. Looking at the Parallel validation accuracy in Table \ref{tab:aug-parallel-val}, there is a clear pattern as four of the feature types improve performance on all five prediction tasks and two feature types improve performance on four out of five prediction tasks, when using data augmentation vs. not. This pattern also holds for the Cascade model validation performance in Table \ref{tab:aug-cascade-val} as three feature types improve on all tasks and the other three feature types improve on four out of five tasks.

\begin{table}[H]
\caption{Data augmentation validation accuracy results with the Parallel model. We observe an improvement for both attributes and misfire using data augmentation across all feature types. The only exceptions being the wavelets performing better on fuel type and waveform performing better on misfire without augmentation.}
\label{tab:aug-parallel-val}
\begin{center}
\begin{tabular}{|c|c|c|c|c|c|c|c|} \hline
\textbf{Model} & \textbf{Feature} & \textbf{Aug} & \textbf{Fuel} & \textbf{Config} & \textbf{Cyl} & \textbf{Turbo} & \textbf{Misfire} \\ \hline
Parallel & FFT & N & 86.7\% & 86.5\% & 73.3\% & 63.5\% & 72.5\% \\ 
Parallel & FFT & Y & \textbf{90.1\%} & \textbf{89.9\%} & \textbf{78.2\%} & \textbf{80.0\%} & \textbf{84.4\%} \\ 
\midrule
Parallel & MFCC & N & 95.0\% & 92.9\% & 87.9\% & 90.8\% & \textbf{89.5\%} \\ 
Parallel & MFCC & Y & \textbf{96.0\%} & \textbf{94.0\%} & \textbf{93.6\%} & \textbf{92.9\%} & \textbf{89.4\%} \\ 
\midrule
Parallel & Spectrogram & N & 86.9\% & \textbf{86.6\%} & 73.4\% & 63.5\% & 82.0\% \\ 
Parallel & Spectrogram & Y & \textbf{87.7\%} & \textbf{86.7\%} & \textbf{75.9\%} & \textbf{80.0\%} & \textbf{85.8\%} \\ 
\midrule
Parallel & Waveform & N & 89.9\% & 72.3\% & 71.4\% & 87.4\% & \textbf{84.2\%} \\ 
Parallel & Waveform & Y & \textbf{95.2\%} & \textbf{93.5\%} & \textbf{90.1\%} & \textbf{90.5\%} & 81.1\% \\ 
\midrule
Parallel & Wavelets & N & \textbf{86.6\%} & 36.2\% & 39.3\% & 73.3\% & 85.7\% \\ 
Parallel & Wavelets & Y & 85.9\% & \textbf{65.4\%} & \textbf{53.8\%} & \textbf{80.6\%} & \textbf{87.5\%} \\ 
\hline
\end{tabular}
\end{center}
\end{table}

\begin{table}[H]
\caption{Data augmentation validation accuracy results with the Cascade model. Similar to the Parallel model, we observe an improvement for both attributes and misfire using data augmentation across all feature types. The only exceptions being the wavelets performing better on fuel type and waveform performing better on misfire without augmentation.}
\label{tab:aug-cascade-val}
\begin{center}
\begin{tabular}{|c|c|c|c|c|c|c|c|} \hline
\textbf{Model} & \textbf{Feature} & \textbf{Aug} & \textbf{Fuel} & \textbf{Config} & \textbf{Cyl} & \textbf{Turbo} & \textbf{Misfire} \\ \hline
Cascade & FFT & N & 86.7\% & 86.6\% & 73.4\% & 63.5\% & 84.3\% \\ 
Cascade & FFT & Y & \textbf{90.4\%} & \textbf{89.9\%} & \textbf{81.4\%} & \textbf{77.5\%} & \textbf{91.8\%} \\ 
\midrule
Cascade & MFCC & N & \textbf{96.1\%} & 92.4\% & 89.0\% & 92.2\% & 92.2\% \\ 
Cascade & MFCC & Y & 95.6\% & \textbf{93.7\%} & \textbf{94.0\%} & \textbf{92.8\%} & \textbf{93.6\%} \\ 
\midrule
Cascade & Spectrogram & N & \textbf{88.1\%} & \textbf{86.6\%} & 73.4\% & 64.4\% & 83.2\% \\
Cascade & Spectrogram & Y & 87.4\% & \textbf{86.6\%} & \textbf{76.2\%} & \textbf{78.4\%} & \textbf{86.5\%} \\ 
\midrule
Cascade & Waveform & N & 90.4\% & 78.4\% & 77.1\% & 86.6\% & 83.0\% \\ 
Cascade & Waveform & Y & \textbf{94.9\%} & \textbf{92.0\%} & \textbf{90.8\%} & \textbf{92.0\%} & \textbf{93.4\%} \\ 
\midrule
Cascade & Wavelets & N & \textbf{86.7\%} & 38.5\% & 40.6\% & 75.8\% & 86.8\% \\
Cascade & Wavelets & Y & \textbf{86.6\%} & \textbf{76.4\%} & \textbf{67.9\%} & \textbf{81.4\%} & \textbf{89.4\%} \\ 
\hline
\end{tabular}
\end{center}
\end{table}

When looking at the test set performance for both our models, another pattern shows strongly the value of data augmentation. With the Parallel test set accuracy in Table \ref{tab:aug-parallel-test}, MFCC improves on 4/5 tasks, and waveform improves on 5/5 tasks. FFT and Wavelets only improve on 2/5 tasks, but experience relatively minor degradation on the other three tasks. Similarly, with spectrogram it only improved on the misfire task but achieved the same performance on 2/5 tasks and minor degradation on the other two tasks.\\

The Cascade model further cements the need for data augmentation in its test set results seen in Table \ref{tab:aug-cascade-test}. FFT and waveform improve on 4/5 tasks, spectrogram and Wavelets improve on 3/5 tasks with minor degradation for the remaining two tasks. MFCCs are the only feature set to arguably perform better without data augmentation than with data augmentation, improving on only 2/5 tasks.\\

In summary, the majority of our feature sets for both Parallel and Cascade models improve on most prediction tasks when using data augmentation. This shows the profound impact training with augmented samples has on acoustic characterization neural networks.

\begin{table}[H]
\caption{Data augmentation test accuracy results with the Parallel model. We find the strong validation performance from the Parallel model does not translate to every feature type and task as augmentation improves performance for some features and tasks but not others. }
\label{tab:aug-parallel-test}
\begin{center}
\begin{tabular}{|c|c|c|c|c|c|c|c|} \hline
\textbf{Model} & \textbf{Feature} & \textbf{Aug} & \textbf{Fuel} & \textbf{Config} & \textbf{Cyl} & \textbf{Turbo} & \textbf{Misfire} \\ \hline
Parallel & FFT & N & \textbf{81.0\%} & \textbf{78.5\%} & \textbf{69.0\%} & 67.2\% & 78.4\% \\ 
Parallel & FFT & Y & 78.8\% & 78.3\% & 67.7\% & \textbf{70.6\%} & \textbf{79.5\%} \\
\midrule
Parallel & MFCC & N & 82.0\% & 75.2\% & 67.7\% & \textbf{77.0\%} & 77.9\% \\ 
Parallel & MFCC & Y & \textbf{84.7\%} & \textbf{78.3\%} & \textbf{72.0\%} & 73.0\% & \textbf{85.3\%} \\ 
\midrule
Parallel & Spectrogram & N & \textbf{81.7\%} & \textbf{78.3\%} & \textbf{68.9\%} & \textbf{66.8\%} & 81.9\% \\ 
Parallel & Spectrogram & Y & 77.0\% & \textbf{78.3\%} & \textbf{68.9\%} & 65.9\% & \textbf{84.9\%} \\ 
\midrule
Parallel & Waveform & N & 71.2\% & 63.2\% & 46.1\% & 62.5\% & 68.7\% \\ 
Parallel & Waveform & Y & \textbf{73.3\%} & \textbf{75.2\%} & \textbf{64.7\%} & \textbf{67.7\%} & \textbf{70.8\%} \\ 
\midrule
Parallel & Wavelets & N & \textbf{81.8\%} & 30.8\% & 24.7\% & \textbf{69.2\%} & \textbf{82.0\%} \\ 
Parallel & Wavelets & Y & 80.1\% & \textbf{63.3\%} & \textbf{40.3\%} & 68.9\% & 81.5\% \\ 
\hline
\end{tabular}
\end{center}
\end{table}

\begin{table}[H]
\caption{Data augmentation test accuracy results with the Cascade model. Similar to the Parallel model, we observe mixed results as augmentation improves performance for some features and tasks but not others. }
\label{tab:aug-cascade-test}
\begin{center}
\begin{tabular}{|c|c|c|c|c|c|c|c|} \hline
\textbf{Model} & \textbf{Feature} & \textbf{Aug} & \textbf{Fuel} & \textbf{Config} & \textbf{Cyl} & \textbf{Turbo} & \textbf{Misfire} \\ \hline
Cascade & FFT & N & \textbf{81.6\%} & \textbf{78.2\%} & \textbf{68.8\%} & \textbf{67.1\%} & 77.1\% \\ 
Cascade & FFT & Y & \textbf{81.7\%} & 77.8\% & 68.1\% & 66.5\% & \textbf{83.5\%} \\
\midrule
Cascade & MFCC & N & 80.0\% & 75.4\% & 66.1\% & \textbf{72.5\%} & 82.4\% \\ 
Cascade & MFCC & Y & \textbf{81.6\%} & \textbf{78.2\%} & \textbf{67.3\%} & 70.1\% & \textbf{86.8\%} \\ 
\midrule
Cascade & Spectrogram & N & 80.9\% & \textbf{78.4\%} & \textbf{68.9\%} & 66.7\% & 82.2\% \\ 
Cascade & Spectrogram & Y & \textbf{81.7\%} & \textbf{78.3\%} & \textbf{68.9\%} & \textbf{67.2\%} & \textbf{85.9\%} \\
\midrule
Cascade & Waveform & N & 76.1\% & 66.6\% & 49.5\% & 63.2\% & 75.4\% \\ 
Cascade & Waveform & Y & \textbf{82.0\%} & \textbf{78.3\%} & \textbf{52.5\%} & \textbf{67.3\%} & \textbf{79.6\%} \\
\midrule
Cascade & Wavelets & N & \textbf{81.7\%} & 29.3\% & 25.3\% & \textbf{68.5\%} & 77.4\% \\ 
Cascade & Wavelets & Y & 78.4\% & \textbf{62.1\%} & \textbf{53.1\%} & 66.8\% & \textbf{83.1\%} \\
\hline
\end{tabular}
\end{center}
\end{table}

\subsection{Feature Fusion}
Feature fusion is a common technique in the building of deep learning systems with which the goal is to create a more comprehensive and robust model by combining multiple feature sets together. In the case of our work, we explore five unique audio feature types: FFT, MFCC, spectrogram, wavelets, and waveform. We can directly observe the value added by specific feature types in Table \ref{tab:feature-test} as the 2D features of MFCC and spectrogram provide the the best insights into the misfire detection whereas the 1D feature of the raw waveform provides the most insight into certain attribute prediction tasks.\\

We face a unique challenge when fusing these audio features together in that each represents a distinct dimensionality. Per Section \ref{sec:network}, the dimensions are as follows: raw waveform and wavelets are  $1\times72,000$, the FFT is $1\times24,000$, the MFCCs is $130\times13$, and the spectrogram is $1025\times282$. For simplicity, we consider one fusion technique: concatenation. To fuse our entire feature set together, we flatten the 2D features of MFCC and spectrogram and then concatenate all features into a long 1D vector.\\

We theorized feature fusion could take the best parts of each feature type and should perform at least as well as its best individual feature set. However, instead we observed concatenation had an overall negative effect on performance. In particular, in Tables \ref{tab:fusion-val} and \ref{tab:fusion-test}, we can see that  feature fusion via concatenation performs worse than the top-performing feature across all attributes and misfire prediction tasks for both validation and test set accuracy.\\

Concatenation appears to settle in between all the features serving as an averaging of sorts. Na{\"i}ve feature fusion through concatenation does not appear to be the correct solution most likely due to the flattening of the 2D features and a smaller relative receptive field with the long 1D input vector. In turn, implementing more complex feature-preserving fusion techniques through multi-lane or ensemble networks is a goal for future work.

\begin{table}[H]
\caption{Feature fusion via concatenation for validation set accuracy.}
\label{tab:fusion-val}
\begin{center}
\begin{tabular}{|c|c|c|c|c|c|c|c|} \hline
\textbf{Model} & \textbf{Feature} & \textbf{Fuel} & \textbf{Config} & \textbf{Cyl} & \textbf{Turbo} & \textbf{Misfire} \\ \hline
Parallel & Top-performing & \textbf{96.0\%} & \textbf{94.0\%} & \textbf{93.6\%} & \textbf{92.9\%} & \textbf{89.4\%} \\
Parallel & Concatenation & 92.1\% & 92.3\% & 82.2\% & 80.8\% & 88.5\% \\ 
\midrule
Cascade & Top-performing & \textbf{95.6\%} & \textbf{93.7\%} & \textbf{94.0\%} & \textbf{92.8\%}  & \textbf{93.6\%} \\ 
Cascade & Concatenation & 91.9\% & 92.7\% & 89.2\% & 75.3\% & 91.2\% \\ 
\hline
\end{tabular}
\end{center}
\end{table}

\begin{table}[H]
\caption{Feature fusion via concatenation for test set accuracy.}
\label{tab:fusion-test}
\begin{center}
\begin{tabular}{|c|c|c|c|c|c|c|} \hline
\textbf{Model} & \textbf{Feature} & \textbf{Fuel} & \textbf{Config} & \textbf{Cyl} & \textbf{Turbo} & \textbf{Misfire} \\ \hline
Parallel & Top-performing & \textbf{84.7\%} & \textbf{78.3\%} & \textbf{72.0\%} & \textbf{73.0\%} & \textbf{85.3\%} \\ 
Parallel & Concatenation & 82.6\% & 77.2\% & 56.0\% & 67.9\% & 82.2\% \\ 
\midrule
Cascade & Top-performing & \textbf{82.0\%} & \textbf{78.3\%} & \textbf{68.9\%} & \textbf{70.1\%} & \textbf{86.8\%} \\ 
Cascade & Concatenation & 81.7\% & 56.9\% & 61.2\% & 68.2\% & 79.4\% \\
\hline
\end{tabular}
\end{center}
\end{table}

\subsection{YouTube Outsample} \label{sec:ablation-YT}
We collected $51$ samples from YouTube (YT) to provide our models with representative data for evaluating outsample generalizability. After chunking these samples every three seconds, we built a new test set of $649$ never-before-seen clips to evaluate the outsample performance of our model.\\

We compare the test set performance of our Parallel and Cascade models to the YT outsample performance in Tables \ref{tab:yt-parallel-test} and \ref{tab:yt-cascade-test}. We observe a significant performance degradation of our models on these YouTube clips. Specifically, in Table \ref{tab:yt-parallel-test} for the Parallel model, we find for FFT, MFCC, and waveform features, there is minimal to extreme degradation in all five prediction tasks. Additionally, Wavelets degrade on 4/5 tasks and Spectrogram, although shown to be the most robust, still degrades on 3 tasks. For example, to show how significant this degradation is when comparing our top performing Parallel model on the test set, we see $11.1\%$, $38.3\%$, $61.3\%$, $4.2\%$, and $22.9\%$ decreases in accuracy across the fuel, config, cyl, turbo, and misfire tasks, respectively.\\

This pattern is also reflected in the Cascade model as seen in Table \ref{tab:yt-cascade-test}. In five of the feature types, all five tasks show degradation and the remaining feature type has degradation in 4/5 tasks. With the top-performing Cascade Spectrogram model, we see $1.3\%$, $41.6\%$, $57.7\%$, $12.8\%$, and $20.0\%$ decreases in accuracy across the fuel, config, cyl, turbo, and misfire tasks, respectively.

\begin{table}[H]
\caption{YouTube outsample compared to traditional test set performance for the Parallel model. We find across all feature types and prediction tasks either a significant degradation or a marginal improvement when comparing the YouTube outsample vs. traditional test set.}
\label{tab:yt-parallel-test}
\begin{center}
\begin{tabular}{|c|c|c|c|c|c|c|c|} \hline
\textbf{Model} & \textbf{Feature} & \textbf{Set} & \textbf{Fuel} & \textbf{Config} & \textbf{Cyl} & \textbf{Turbo} & \textbf{Misfire} \\ \hline
Parallel & FFT & Test & \textbf{78.8\%} & \textbf{78.3\%} & \textbf{67.7\%} & \textbf{70.6\%} & \textbf{79.5\%} \\ 
Parallel & FFT & YT & 75.7\% & 36.6\% & 9.9\% & 49.6\% & 65.4\% \\
\midrule
Parallel & MFCC & Test & \textbf{84.7\%} & \textbf{78.3\%} & \textbf{72.0\%} & \textbf{73.0\%} & \textbf{85.3\%} \\ 
Parallel & MFCC & YT & 73.6\% & 40.0\% & 10.7\% & 68.8\% & 62.4\% \\
\midrule
Parallel & Spectrogram & Test & 77.0\% & \textbf{78.3\%} & \textbf{68.9\%} & 65.9\% & \textbf{84.9\%} \\ 
Parallel & Spectrogram & YT & \textbf{78.7\%} & 36.9\% & 11.5\% & \textbf{67.9\%} & 63.3\% \\
\midrule
Parallel & Waveform & Test & \textbf{73.3\%} & \textbf{75.2\%} & \textbf{64.7\%} & \textbf{67.7\%} & \textbf{70.8\%} \\ 
Parallel & Waveform & YT & 58.9\% & 39.3\% & 22.2\% & 49.4\% & 64.7\% \\
\midrule
Parallel & Wavelets & Test & \textbf{80.1\%} & \textbf{63.3\%} & 40.3\% & \textbf{68.9\%} & \textbf{81.5\%} \\ 
Parallel & Wavelets & YT & 73.5\% & 45.0\% & \textbf{41.2\%} & 52.8\% & 54.0\% \\
\hline
\end{tabular}
\end{center}
\end{table}

\begin{table}[H]
\caption{YouTube outsample compared to traditional test set performance for the Cascade model. Similar to the Parallel model, we find across all feature types and prediction tasks either a significant degradation or a marginal improvement when comparing the YouTube outsample vs. traditional test set.}\label{tab:yt-cascade-test}
\begin{center}
\begin{tabular}{|c|c|c|c|c|c|c|c|} \hline
\textbf{Model} & \textbf{Feature} & \textbf{Set} & \textbf{Fuel} & \textbf{Config} & \textbf{Cyl} & \textbf{Turbo} & \textbf{Misfire} \\ \hline
Cascade & FFT & Test & \textbf{81.7\%} & \textbf{77.8\%} & \textbf{68.1\%} & \textbf{66.5\%} & \textbf{83.5\%} \\
Cascade & FFT & YT & 77.0\% & 35.2\% & 11.2\% & 45.7\% & 67.3\% \\ 
\midrule
Cascade & MFCC & Test & \textbf{81.6\%} & \textbf{78.2\%} & \textbf{67.3\%} & \textbf{70.1\%} & \textbf{86.8\%} \\ 
Cascade & MFCC & YT & 76.5\% & 42.5\% & 18.4\% & 68.0\% & 64.4\% \\ 
\midrule
Cascade & Spectrogram & Test & \textbf{81.7\%} & \textbf{78.3\%} & \textbf{68.9\%} & \textbf{67.2\%} & \textbf{85.9\%} \\
Cascade & Spectrogram & YT & 80.6\% & 36.7\% & 11.4\% & 54.6\% & 63.2\% \\ 
\midrule
Cascade & Waveform & Test & \textbf{82.0\%} & \textbf{78.3\%} & \textbf{52.5\%} & \textbf{67.3\%} & \textbf{79.6\%} \\
Cascade & Waveform & YT & 71.6\% & 41.6\% & 24.1\% & 52.5\% & 62.0\% \\
\midrule
Cascade & Wavelets & Test & \textbf{78.4\%} & \textbf{62.1\%} &\textbf{53.1\%} & \textbf{66.8\%} & \textbf{83.1\%} \\
Cascade & Wavelets & YT & 72.1\% & 43.6\% & 35.2\% & 54.2\% & 63.6\% \\
\hline
\end{tabular}
\end{center}
\end{table}

\begin{figure}[H]
\begin{center}
 \caption{Example FFT - YT Sample - Frequency attenuated (left) and regular sample (right)}
 \includegraphics[width=0.45\textwidth]{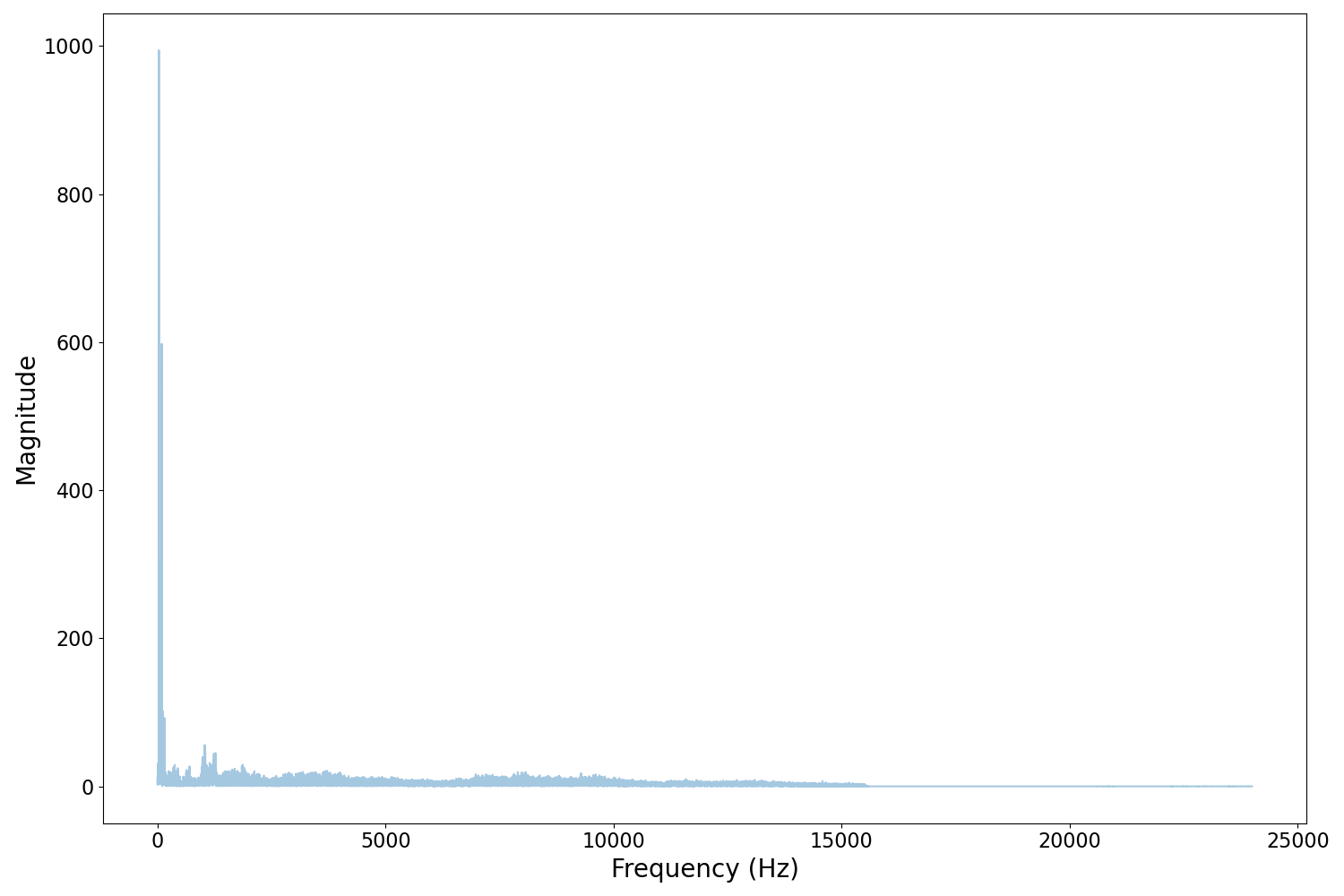}
 \includegraphics[width=0.45\textwidth]{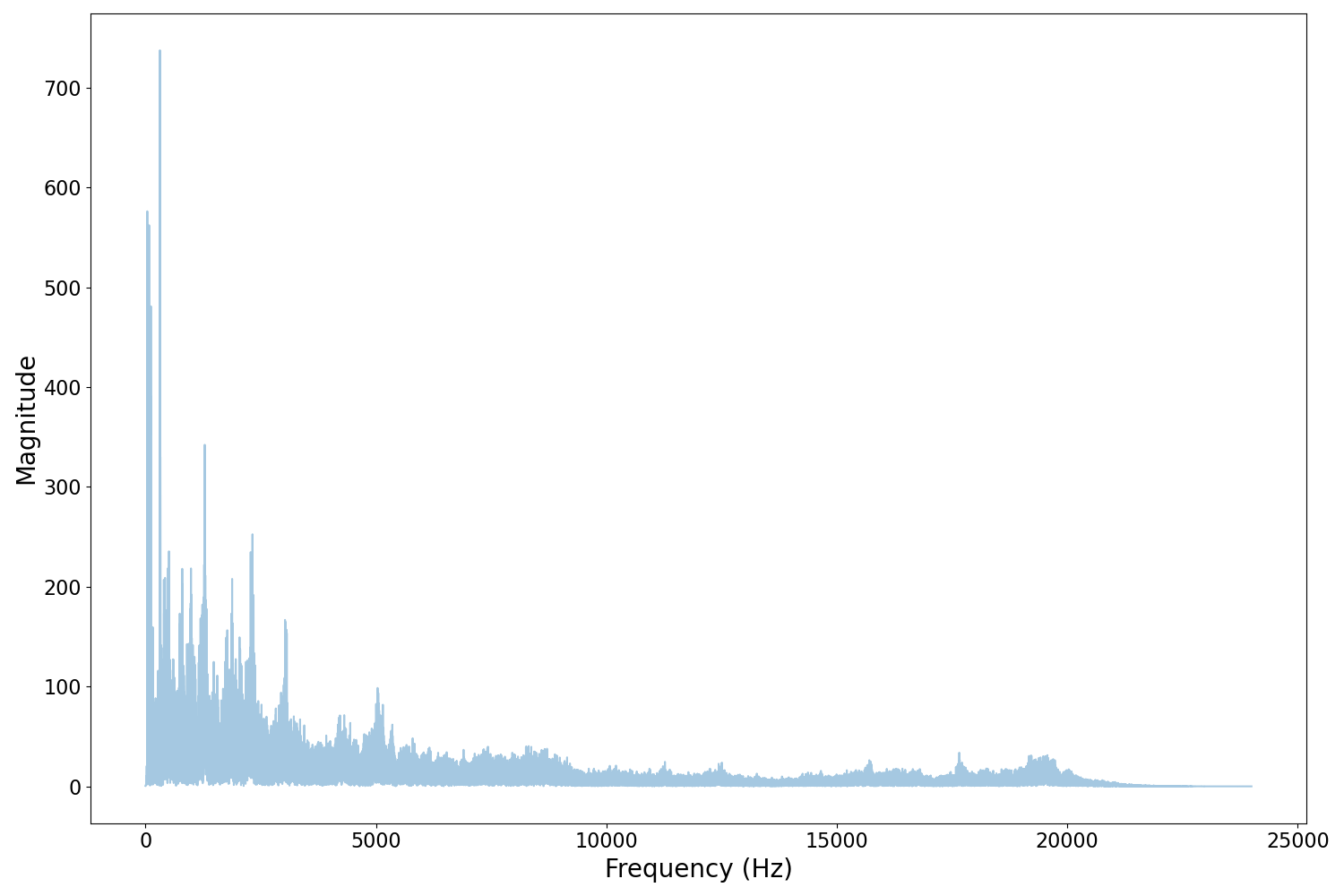}
 \label{fig:fft-freq-atten}
 \end{center}
\end{figure}

\begin{figure}[H]
\begin{center}
 \caption{Example Spectrogram - YT Sample - Frequency attenuated (left) and regular sample (right)}
 \includegraphics[width=0.45\textwidth]{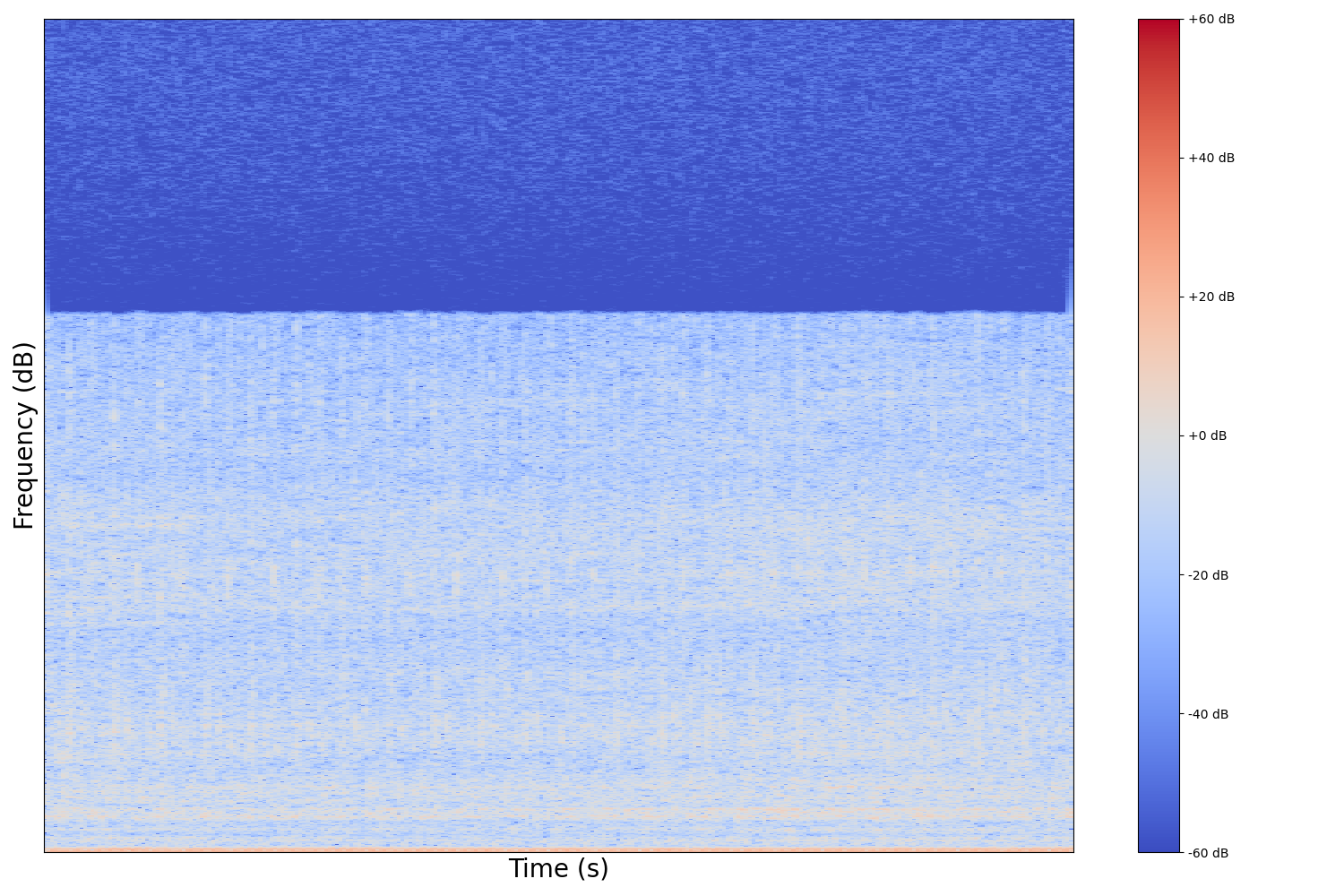}
 \includegraphics[width=0.45\textwidth]{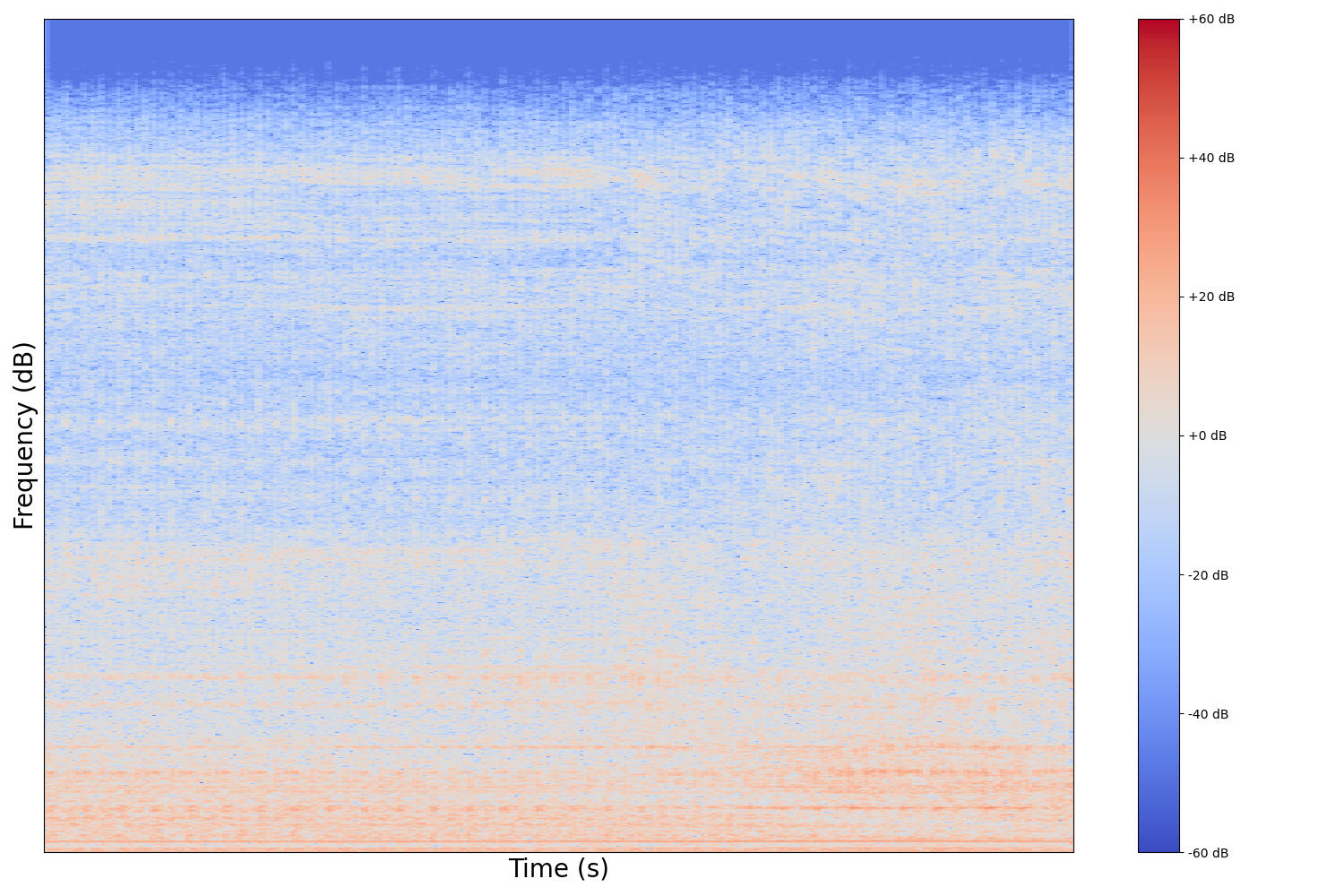}
 \label{fig:spec-freq-atten}
 \end{center}
\end{figure}

There are a handful of reasons why this sort of discrepancy has occurred between our test set and the YouTube test set. It could have resulted from a lack of sufficient data and a label distribution mismatch compared to the training set. However, we have evidence that YouTube audio is frequency-attenuated on certain videos, likely to save space. We show an example of a regular sample from our set and a sample from the YouTube set in Figures \ref{fig:fft-freq-atten} and \ref{fig:spec-freq-atten}. In Figure~\ref{fig:fft-freq-atten}, note that there is no representation of frequencies greater than $16$~kHz in the YT sample whereas in raw device-captured samples we see frequencies represented well beyond $20$~kHz. Given our sampling rate is $48$~kHz, according to the Nyquist-Shannon Theorem, our features should have informative frequency representation up to $24$~kHz.\\ \\

From a physics standpoint, some features, such as turbocharger bearing rotation, would make acoustic emissions around these higher frequencies. The comparison of Spectrograms shown in  Figure~\ref{fig:spec-freq-atten} further elucidates the frequency attentuation. For the Youtube sample, in  Figure~\ref{fig:spec-freq-atten}, we can visualize a sharp knee in the function at a much lower frequency than appears in the regular sample Spectrogram.\\

In summary, we showed that there should be caution applied when working with crowd-sourced acoustic data such as that in datasets originating on YouTube. This further motives the need for more comprehensive data collection and generation to create a balanced and diverse set for testing the outsample performance of models.

\section{Discussion}
In this section we will explore three main areas of discussion:
\begin{itemize}
 \item Broader Implications
 \item Future Directions
 \item Potential Applications
\end{itemize}
Within broader implications, we consider the larger impact our proposed architecture may have on both consumers and industry. For future directions, we note routes this work could be continued and built upon with respect to data quality, model improvement, and architectural modifications. Finally, we discuss the how our work could be utilized in diverse application areas within and outside vehicle characterization and diagnostics. 

\subsection{Broader Implications}
There are two main groups to consider for the broader impacts of our work developing an AI mechanic: consumers and industry.\\

Given the end-to-end nature of our proposed system, it has practical usability for consumers to obtain vehicle status in real-time, e.g. using a mobile phone. Our high-level cascading architecture takes as input a raw audio sample, requiring no direct input or knowledge from the consumer. This has the impact of potentially saving even non-expert users stress, time, and money. Whenever their personal vehicle may be exhibiting a fault, they can receive an initial opinion from the AI mechanic. This AI mechanic could provide the consumer with insight into the seriousness of their vehicle state.\\

There are multiple extensions and improvements the AI mechanic may uncover when consumer data is integrated. This would create a feedback loop that would only improve the system at becoming better at detecting faults. Additionally, collection of fault data over time from long-standing consumers would create the opportunity for preventive maintenance prediction and remaining useful life (RUL) prediction. The greatest impact the AI mechanic could have on the average consumer would be helping to prevent their vehicle from exhibiting a fault in the first place through providing warnings before a fault occurs.\\

Prognostics will not come about immediately. For there to be value in a feedback loop from consumers, vehicles must be allowed to fail such that these data might be captured. Additionally, we showed the impact underrepresented labels have on our model and its poor outsample performance on most underrepresented labels. Without a more diverse, comprehensive, and well-balanced dataset, this behavior will still be observed. When the AI mechanic is working with a previously-unseen vehicle, it will take the system time to adjust, and may necessitate the use of techniques such as synthetic data generation or federated learning.\\

Within the automotive industry, off-board diagnostics which can utilize inexpensive microphones and mobile or embedded devices an alternative to the costly expense of on-board diagnostics that constantly collect and send data from the vehicle. The AI mechanic could provide automakers with an enormous swath of data on the status condition and fault identification for each kind of vehicle.\\

Our cascading models are flexible in that despite being trained using the vehicle attribute predictions as input to the second stage network, this could easily be adapted to take the ground truth labels as input instead. The value of this change could be significant for automakers as it would allow them to build unique fault identification models conditional upon specific vehicle types. If automakers could use inexpensive audio data to create a chart of anticipated vehicle wear, this may be used to design better vehicles and supporting infrastructure. This will require the capture and labeling of samples for desired fault identification types. If other fault types were to be integrated into the AI mechanic, this would require hundreds or thousands of distinct examples across vehicle classes to achieve similar performance to our method's demonstrated misfire detection performance. 

\subsection{Future Directions}
Throughout our results and ablation sections, we brought up issues with our current approach and raised potential opportunities for improvement. One of the main issues uncovered (and not unique to the AI mechanic) is the models' dependency on quality data that is not only diverse but also balanced across classes. We showed that obtaining crowd-sourced audio data from YouTube may address data quantity, but not data quality. There is value in developing a real-world application that would make it easy for a consumer to collect data on their mobile device and store the waveform at its original $48$~kHz sampling rate to prevent lossy compression or frequency attenuation.\\

Another avenue for future work to address the lack of data, especially in underrepresented classes, would be generative modeling. In particular, Generative Adversarial Networks (GANs)~\cite{goodfellow2014generative} have shown great power in developing realistic samples in numerous fields, most notably facial recognition. Further exploration would be needed to understand how GANs could be extended not only to acoustic samples, but also utilized in a feedback loop for developing better synthetic samples for training, including for classes not well represented.\\

Another direction for improving validation and test set accuracy would be developing a better approach for feature and model fusion. Our attempts at feature fusion through concatenation by flattening of the informative 2D features were in vain. Likely a multi-lane or ensemble approach where 1D and 2D convolution can be utilized on their respective feature types should yield better overall performance. Additionally, we may consider mixup \cite{zhang2018mixup, xu2018mixup} data augmentation as a way to improve training and generate more robust features.\\

There is a significant opportunity in the space of acoustic vehicle characterization. Our dataset had labels for make, OEM, idling status, recording environment/location, and more. Perhaps there could be value derived from more expansive attribute classification. Additionally, our set contains labels for horsepower and engine displacement, which would provide an interesting exploration in attribute regression.\\

Finally, building upon our initial proof-of-concept for a cascading architecture would be a challenging but worthwhile endeavor. It could be tackled in two main ways. The first way would be to connect more stages to the architecture both in the high-level, general acoustic classification, and low-level, fault type recognition. These additional stages may benefit from utilizing pre-trained audio embeddings and transfer learning~\cite{arandjelovic2017look, cramer2019look, kong2020panns, gong2021ast}. Additionally, extensions of existing layers may be explored to increase the level to which a system is characterized such as RUL if the status is detected as normal. The other approach, motivated by work discussed in Section~\ref{related:cascade}, would be to implement conditional logic into the network itself. Since fault type recognition is conditional upon there being a fault in the first place, having a network which is adaptive would be a major contribution. There has already been great strides made in deep sequential networks~\cite{denoyer2014deep} and adaptive networks focused on efficiency~\cite{Wu_2018_CVPR, shazeer2017outrageously, bengio2015conditional}. As such, building this work upon prior art would be a natural extension going forward.

\subsection{Potential Applications}
While our main focus of this work was on vehicle understanding, our approach and proposed cascading architecture can be adapted in  broader application areas. These application areas include sequential, conditional, or multi-label prediction tasks and fault identification.\\

Our proposed cascading architecture is a multi-level network where each level is conditional upon the prior. Any application area where there is potential for inter-label dependency could leverage the cascading architecture. For example, music recognition has hierarchies and label dependency such as first asking does the sample contain music. Then the next stage is conditional upon the first stage where it could look to predict genre, artist, song, etc. Another area for cascading architecture could be animal recognition. Does the sample first contain an animal, then what kind of animal, what state is the animal in, what location is the animal in, is the animal behaving normally, etc.?\\

Not only can the cascading architecture be extended to other audio applications, but also applications with other modes of data such as in computer vision. For example, biometrics could first ascertain whether a sample contains a valid fingerprint, iris, or facial scan. Then conditionally upon the first level, it could then ask whether the biometrics scan represents a valid user, what condition the user is in, perhaps using multi-modal data such as heart rate or blood pressure prediction. Another particularly relevant example in the larger vision field is autonomous vehicles (AVs). These systems are fusing many modes of data from sensors and making certain the state of the AV would be crucial. Again it could follow the hierarchy of does the sample contain an AV, what are the attributes of the AV, is the AV behaving normally, and if not what fault behavior is the AV?\\

The other significant application area for these techniques is broader fault identification, particularly using audio data. This includes other diagnostic areas, such as for industrial processes or energy sector equipment. One such example is home or industrial heating and ventilation systems: in this case, we first ask can we identify whether a sample contains ventilation equipment using acoustic classification networks. Then we would look to obtain its operating state and condition. If it's behaving normally, what is expected RUL? If abnormal, what is the fault type and degree? \\

This cascading architecture can be extended to any application where a fault might occur. Some of these may include home appliances (washer/dryer) with belt slipping or drum imbalance, electric cars / bicycles with suspension issues, manufacturing equipment (CNC mills/lathes) with tool run-out or spindle issues, drills with brush wear or belt slip, the energy sector with turbine and pump health, elevator / escalators condition, and even carnival / fair equipment.\\

In summary, our proposed method is broadly applicable to any task which can leverage a conditional dependency and/or has a potential need for fault identification. We plan to explore the identified areas and more in continuations of this work. 
\newpage
\section{Conclusion} 
In this manuscript, we took the first steps towards building an AI mechanic. We identified four main areas of novelty with our work:
\begin{itemize}
 \item \textbf{Sound Recognition using Deep Learning}: We demonstrated the value and need for vehicle understanding as an application domain for sound recognition using deep learning. More specifically, we know the world is becoming ever dependent upon transportation. Sound is a prevalent and inexpensive resource. Fault detection is a lengthy and time-consuming endeavor even for expert mechanics. Through an AI mechanic that can listen to a vehicle, we can utilize a cost-effective solution that can improve lives by reducing vehicle emissions and increasing usable service life.\\
 
 By developing knowledge and capabilities in support of identifying systems with expert level detail, without expert knowledge, we reduce barriers to instrumentation that today stand in the way of proactive and reactive maintenance operations. Our solution will help individuals, fleet managers, and third parties to instrument vehicles more comprehensively and effectively than is presently feasible. The net result of this increased insight and transparency will be a more efficient, reliable, and safer fleet.\\
 
 \item \textbf{Acoustic Vehicle Characterization}: We showcased the first published results on the engine configuration attribute task, to the best of our knowledge. Improving on prior work, we built an all-in-one model for all four attribute types. Additionally, we employed multi-task learning for both the attributes and misfire prediction tasks to better inform one another. We also merged the attributes and misfire datasets which created a larger and more challenging dataset to predict attributes on both normally and abnormally performing samples.\\
 
 \item \textbf{Cascading Architectures}: We defined cascading architectures as sequential, multi-level, conditional networks. As a proof-of-concept, we built a two-stage convolutional neural network where the first stage specialized on vehicle attributes and then cascaded its attribute predictions to the second stage which specialized on misfire detection with the goal of better informing both prediction tasks.\\
 
 This two-stage cascading network demonstrated strong performance of $95.6\%$, $93.7\%$, $94.0\%$, and $92.8\%$ validation accuracy on the fuel type, engine configuration, cylinder count, and aspiration type prediction tasks, respectively. In comparing across a na{\"i}ve and parallel baseline, the cascading model showcased $16.4\%$ and $4.2\%$, respectively, relative improvement on misfire fault prediction with a validation accuracy of $93.6\%$. Additionally, the cascading model achieved $86.8\%$ test set accuracy on misfire fault prediction, demonstrating margins of $7.8\%$ and $1.5\%$ improvement over na{\"i}ve and parallel baselines.\\
 
 \item \textbf{Audio Data Augmentation}: We constructed a vast acoustic vehicle characterization dataset with over 50k samples for over 40 hours of audio using data augmentation techniques. Through ablation studies, we establish the positive impact data augmentation has on outsample generalizability. We additionally show the limitations of crowd-sourced audio from YouTube to further support the value our augmented dataset provides to the community.\\ 
\end{itemize} 

We concluded our manuscript with an important discussion of the broader implications, future directions, and potential applications of the AI mechanic, highlighting the potential for future research and adoption of the developed architectures and concepts.
\newpage
\bibliographystyle{plain}
\bibliography{main.bbl}

\end{document}